# Correcting Bartlett factor-score error for latent covariates in partially linear double machine learning

Eunjeong Song · Sehee Hong
Department of Education, Korea University, Seoul, Republic of Korea

Eunjeong Song, ORCID iD: https://orcid.org/0000-0002-2302-3227
Sehee Hong, ORCID iD: https://orcid.org/0000-0001-5468-8398

**Abstract**

Substituting Bartlett factor scores for latent covariates leaves factor-score error that does not vanish as the sample grows and biases the adjusted structural coefficient. We propose Croon-DML, which subtracts the factor-score error covariance implied by the measurement model from the cross-fitted estimating equation of partially linear double/debiased machine learning (DML), and establish identification, Neyman orthogonality with respect to the conditional means, consistency, and asymptotic normality. Residual bias after correction comes from learning the conditional means of the factor scores: signal learning error, and noise fitting, in which the learner also fits the factor-score error. Signal learning error largely cancels when the three conditional means share a common penalty, and noise fitting by a linear smoother can be computed from its weights alone, so we also subtract it (Croon-DML-R). Across 12 Monte Carlo conditions, Croon-DML reduced the absolute bias of uncorrected DML by 66–95%, and the bias of Croon-DML-R with a common penalty differed from that of an oracle estimator using the true latent variables by at most 0.004. In an illustration with public longitudinal data, the correction was 15% of the latent-covariate adjustment; its size reflected the latent-covariate coefficients rather than the residualized reliabilities.

## 1. Introduction

Observational studies in psychology and education typically estimate the relationship between a focal predictor and an outcome after adjusting for covariates. A study of study time and mathematics identity, for example, might adjust for gender, prior achievement, and latent covariates such as prior identity and self-efficacy, each measured by multiple indicators. Latent covariates are commonly replaced by factor scores and treated as observed (Grice, 2001; DiStefano et al., 2009), but factor scores differ from the latent variables (Rhemtulla & Savalei, 2025). With a fixed number of indicators of fixed quality, this discrepancy does not vanish as the sample grows, and it biases the adjusted coefficient (Fuller, 1987; Steiner et al., 2011; Cole & Preacher, 2014; Lockwood & McCaffrey, 2016; Westfall & Yarkoni, 2016); with several covariates, biases transmitted along different paths can offset or amplify one another (Sengewald & Pohl, 2019; Sengewald et al., 2019).

Two lines of work address this bias. The first follows Croon (2002) and corrects factor score regression (Skrondal & Laake, 2001; Hoshino & Bentler, 2013) by adjusting the covariances of the scores with information from the measurement model. It has been extended in several directions (Devlieger et al., 2016; Devlieger & Rosseel, 2017; Hayes & Usami, 2020; Rosseel & Loh, 2024; Rosseel et al., 2025; Bogaert et al., 2026), its stepwise estimators have standard errors and consistency results (Can & Rosseel, 2025; Levy, 2026; Liu et al., 2026), and the inclusive (posterior-mean) factor score (Nguyen & Stuart, 2020) belongs to it. These methods model the observed covariates parametrically, usually linearly. The second line—partially linear models (Robinson, 1988) and double/debiased machine learning (DML; Chernozhukov et al., 2018)—estimates unknown functions of many observed covariates by machine learning and removes the first-order effect of learning error through orthogonalized estimating equations and cross-fitting, but assumes covariates observed without error. Work bridging the two lines covers partially linear errors-in-variables models with known error covariance (Liang et al., 1999), mismeasured high-dimensional controls (Kim & Song, 2024), generated regressors (Pagan, 1984; Zhang et al., 2026; Escanciano & Pérez-Izquierdo, 2026), latent-variable DML (Morimura et al., 2025), and effect restoration through proxies (Kuroki & Pearl, 2014). Factor-score error, however, does not vanish with the sample size, so cross-fitting cannot remove it. Learning the conditional means of noisy scores creates a bias of its own. To our knowledge, no study has combined the error covariance from a measurement model with cross-fitted estimation or examined this second bias.

This article makes two contributions. First, Croon-DML subtracts the error covariance obtained from the measurement model from the cross-fitted estimating equation; we show that it identifies the structural coefficient, retains Neyman orthogonality with respect to the conditional means (its expectation is insensitive, to first order, to their learning error), and is consistent and asymptotically normal. Second, we show that learning the conditional mean of a noisy factor score creates a second bias, noise fitting, in which the learner also fits the factor-score error; for a linear smoother its size follows from the smoother weights, so it can be removed (Croon-DML-R), whereas the learning error in the signal largely cancels when the same smoother is used for all three conditional means.

## 2. Background and definitions

### *2.1 Structural model and conditional means*

For individual $i$, let $Y_i$ be a continuous outcome, $D_i$ the focal predictor, $Z_i$ a vector of observed covariates, $\eta_i$ a $J$-dimensional vector of latent covariates, and $X_i$ the vector of their indicators; $Y_i, D_i, Z_i$ are observed without measurement error (Figure 1; notation in Table 1). The structural model is

$$D = h_{D0}(Z) + \eta'\gamma_0 + v, \qquad Y = \theta_0 D + h_{Y0}(Z) + \eta'\beta_0 + e, \qquad (1)$$

where $h_{D0}$ and $h_{Y0}$ are unknown functions of the observed covariates, $\theta_0$ is the structural coefficient of interest, $\gamma_0$ and $\beta_0$ are the coefficient vectors of the latent covariates in the focal-predictor and outcome equations, and the structural errors satisfy $E(e \mid D, \eta, Z) = 0$ and $E(v \mid \eta, Z) = 0$. If both $\gamma_0$ and $\beta_0$ are nonzero, the latent covariates are confounders.

In DML, the three conditional means

$$\ell_0(Z) = E(Y \mid Z), \qquad m_0(Z) = E(D \mid Z), \qquad q_0(Z) = E(\eta \mid Z) \qquad (2)$$

are called nuisance functions, and the machine learning methods that estimate them are called learners. Subtracting its conditional mean from a variable is called residualization. With $Y^\circ = Y - \ell_0(Z)$, $D^\circ = D - m_0(Z)$, and $\eta^\circ = \eta - q_0(Z)$, the structural model takes the partially linear form

$$Y^\circ = \theta_0 D^\circ + \eta^{\circ\prime}\beta_0 + e, \qquad (1)$$

$$D^\circ = \eta^{\circ\prime}\gamma_0 + v. \qquad (2)$$

The observed covariates, absorbed into $\ell_0$ and $m_0$, may enter nonlinearly or in high dimension; the latent covariates enter linearly after residualization, and their residualized covariance $\Sigma_\eta = \mathrm{Var}(\eta^\circ)$ differs from the factor covariance of the measurement model.

### *2.2 Measurement model and Bartlett factor scores*

We distinguish measurement error in the indicators, factor-score error in the scores, and estimation error in the measurement parameters. The $J$-factor measurement model for $p$ continuous indicators is

$$X = \tau + \Lambda\eta + \varepsilon, \quad E(\varepsilon) = 0, \quad E(\eta) = 0, \quad \mathrm{Cov}(\eta, \varepsilon) = 0, \quad \mathrm{Var}(\varepsilon) = \Theta, \quad \mathrm{Var}(\eta) = \Phi. \qquad (3)$$

$\Lambda$ has linearly independent columns and $\Theta$ is positive definite; the latent means are zero, and each latent variable is scaled by fixing a reference loading or the factor variance to 1 (Bollen, 1989). The factor-score coefficient matrix of Bartlett (1937),

$$B = (\Lambda'\Theta^{-1}\Lambda)^{-1}\Lambda'\Theta^{-1}, \qquad (4)$$

satisfies $B\Lambda = I_J$, so the Bartlett factor scores $S$ can be written as

$$S = B(X - \tau) = \eta + U, \qquad U = B\varepsilon, \qquad (5)$$

and the factor-score error covariance is

$$\Omega = \mathrm{Var}(U) = B\Theta B' = (\Lambda'\Theta^{-1}\Lambda)^{-1}. \qquad (6)$$

We use Bartlett scores because their error is additive, as in equation (5), which suits explanatory latent variables (Skrondal & Laake, 2001), and because $B$ and $\Omega$ do not depend on $\Phi$, so the same values apply before and after residualization; regression factor scores have error correlated with the latent variable and need a different correction (Croon, 2002; Bogaert et al., 2026). If $\Theta$ contains indicator error covariances, $\Omega$ is nondiagonal (Hayes & Usami, 2020).

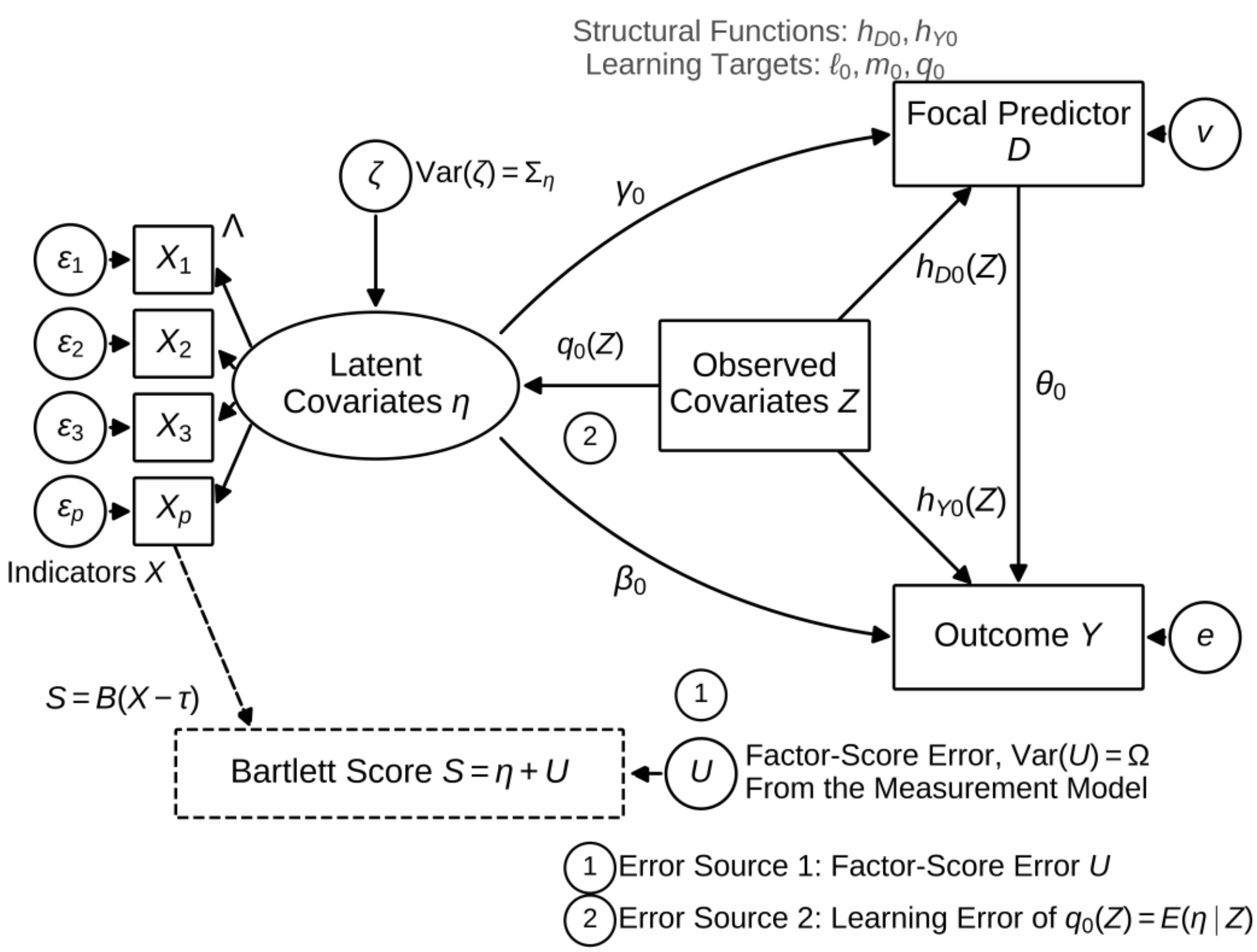

**Figure 1.** The model and its two sources of error. Ellipses denote latent variables, rectangles observed variables, small circles error terms, and single-headed arrows regression relations. The structural functions of $Z$ are $h_{D0}, h_{Y0}, q_0$, and the learning targets are the conditional means

$\ell_0, m_0, q_0$ ($q_0$ is both a structural function and a learning target). $\zeta = \eta^\circ$ is the structural error of the latent covariates. Dashed lines mark the Bartlett score $S = \eta + U$ used in estimation and its error. ① is the factor-score error $U$ and ② is the learning error in $q_0$; $\Omega = \mathrm{Var}(U)$ is computed from the measurement model.

### *2.3 Assumptions*

**A1. Correctly specified and identified measurement model.** The theory-based specification of the loadings, factor covariances, and indicator error covariances matches the data-generating model, and the scale and direction of each factor are fixed consistently, so that $B$ and $\Omega$ are computed correctly.

**A2. Conditional mean and cross-moment conditions for measurement error.** $E(\varepsilon \mid \eta, Z) = 0$, $E(\varepsilon e \mid \eta, Z) = 0$, and $E(\varepsilon v \mid \eta, Z) = 0$. The first condition, that the observed covariates affect the indicators only through the factors, extends measurement invariance across groups (Meredith, 1993; Vandenberg & Lance, 2000) to all of $Z$. It can be relaxed: an intercept shift that is a function of $Z$ is absorbed by residualization, provided that $\Omega$ is estimated from a model that allows it (Section S5.3 of the Supporting Information); loading noninvariance is not absorbed. The other two conditions can fail because of common method effects (Podsakoff et al., 2003) or when the indicators and the outcome share items. A2 and equation (5) imply $E(U \mid \eta, Z) = 0$, $E(Ue) = 0$, and $E(Uv) = 0$, and because $E(S \mid Z) = q_0(Z)$,

$$S^\circ = S - q_0(Z) = \eta^\circ + U, \qquad (7)$$

so the covariance of the residualized Bartlett scores is $\Sigma_\eta + \Omega$. The correction of Croon (2002) then amounts to subtracting $\Omega$ to recover $\Sigma_\eta$.

**A3. Variance conditions.** $\Sigma_\eta$ is positive definite and $\mathrm{Var}(v) > 0$.

### *2.4 Terminology*

*Cross-fitting* splits the sample into $K$ folds and, when evaluating one fold, learns from the remaining folds. The *oracle* uses the true $\eta$ as if observed. The proportion of latent variance in the residualized factor scores, $\mathrm{diag}(\Sigma_\eta)/\mathrm{diag}(\Sigma_\eta + \Omega)$, is the *residualized reliability*. Under the same residualization, the *latent-covariate adjustment* is the difference between the coefficients estimated without and with the latent covariates, and the *factor-score error correction* the difference between the uncorrected and corrected estimates.

**Table 1.** Notation and definitions

| Symbol | Definition |
|---|---|
| $Y, D, Z$ | Outcome, focal predictor, vector of observed covariates (no measurement error) |
| $\eta, X$ | $J$-dimensional latent covariates; vector of their indicators |
| $\theta_0, \gamma_0, \beta_0$ | Structural coefficient of interest; coefficient vectors of the latent covariates in the focal-predictor and outcome equations |
| $h_{D0}, h_{Y0}; v, e$ | Functions of the observed covariates (nonlinear and high-dimensional functions allowed); structural errors |
| $\ell_0, m_0, q_0$ | Conditional means of $Y, D, \eta$ given $Z$ (nuisance functions) |
| $Y^\circ, D^\circ, \eta^\circ, S^\circ$ | Residualized variables (conditional means subtracted); $\zeta = \eta^\circ$ in Figure 1 |

| Symbol | Definition |
|---|---|
| $\tau, \Lambda, \Theta, \Phi$ | Intercepts, factor loadings, error covariance, and factor covariance of the measurement model |
| $B$ | Bartlett factor-score coefficient matrix |
| $S = \eta + U, \Omega$ | Bartlett factor scores; factor-score error $U = B\varepsilon$ and its covariance $\mathrm{Var}(U)$ |
| $\Sigma_\eta, C_{DD}$ | Covariance of the residualized latent covariates, $\mathrm{Var}(\eta^\circ)$; $\mathrm{Var}(D^\circ) = \gamma_0'\Sigma_\eta\gamma_0 + \mathrm{Var}(v)$ |
| $w(Z), \bar{\omega}, G_i$ | Weights of the linear smoother; mean sum of squared weights; matrix of weight inner products for held-out row $i$, $(G_i)_{jk} = w_j(Z_i)'w_k(Z_i)$ |
| $\delta_q, \Delta_q$ | Learning error in $q_0$, $\hat{q} - q_0$; its second moment |
| $\Delta_\eta, \Delta_U, \Delta_U^*$ | Second moments of signal learning error and of noise fitting; approximation to noise fitting (equation 14) |

## 3. Croon-DML

### *3.1 Bias of uncorrected factor-score DML*

Let $W^\circ = (D^\circ, S^{\circ\prime})'$ and $a_0 = (\theta_0, \beta_0')'$. In the population, uncorrected factor-score DML (hereafter uncorrected DML), which treats the residualized factor scores as error-free, solves $E\{W^\circ(Y^\circ - W^{\circ\prime}a)\} = 0$. At the true parameters, $Y^\circ - W^{\circ\prime}a_0 = e - U'\beta_0$, so under A2 and the conditions on the structural errors,

$$E\{W^\circ(Y^\circ - W^{\circ\prime}a_0)\} = \begin{pmatrix} 0 \\ -\Omega\beta_0 \end{pmatrix}. \qquad (8)$$

The latent-variable block thus has expectation $-\Omega\beta_0$ rather than zero, and the discrepancy is transmitted through $\gamma_0$ to $\theta_0$. The bias of the uncorrected structural coefficient $\theta_\mathrm{U}$ is

$$\theta_\mathrm{U} - \theta_0 = \frac{\gamma_0'\Sigma_\eta(\Sigma_\eta + \Omega)^{-1}\Omega\beta_0}{C_{DD} - \gamma_0'\Sigma_\eta(\Sigma_\eta + \Omega)^{-1}\Sigma_\eta\gamma_0} \qquad (9)$$

(Section S1.1). The bias vanishes if $\Omega = 0$, $\beta_0 = 0$, or $\gamma_0 = 0$; otherwise it depends jointly on $\Sigma_\eta, \Omega, \beta_0, \gamma_0$, so reliability alone does not determine it, and even with high reliability the correction need not be small (Figure S1).

### *3.2 Corrected estimating equation and identification*

Let $\Omega_W = \mathrm{blockdiag}(0, \Omega)$, where the zero corresponds to the error-free focal predictor. The Croon-DML estimating equation is

$$g(a; \ell_0, m_0, q_0, \Omega) = W^\circ(Y^\circ - W^{\circ\prime}a) + \Omega_W a. \qquad (10)$$

The individual factor scores are unchanged; the error covariance, here obtained from the measurement model, is subtracted from the second-moment matrix, as in moment-corrected estimators for measurement error models with known error covariance (Fuller, 1987; Liang et al., 1999; Carroll et al., 2006).

**Theorem 1 (Identification).** If A1–A3 and equations (1)–(2) hold, then $A_0 = E(W^\circ W^{\circ\prime}) - \Omega_W$ is positive definite, and $a_0$ is the unique solution of $E\{g(a)\} = 0$,

$$a_0 = [E(W^\circ W^{\circ\prime}) - \Omega_W]^{-1}E(W^\circ Y^\circ). \qquad (11)$$

$A_0$ is the covariance matrix of $(D^\circ, \eta^{\circ\prime})'$, positive definite under A3, and the correction term cancels $-\Omega\beta_0$ in equation (8) (Section S1.2).

### *3.3 Cross-fitted estimation procedure*

Table 2 summarizes the procedure. For each held-out fold $I_k$, the prespecified measurement model is fitted by normal-theory maximum likelihood (ML; Jöreskog, 1969) in the training folds $I_k^c$, giving $\hat{\tau}_k, \hat{B}_k, \hat{\Omega}_k$; the scores $\hat{S}_{ik} = \hat{B}_k(X_i - \hat{\tau}_k)$ and the conditional means $\hat{\ell}_k, \hat{m}_k, \hat{q}_k$ are estimated there, and the residuals $\hat{Y}_i^\circ, \hat{D}_i^\circ, \hat{S}_i^\circ$ are formed in $I_k$. Pooling all held-out folds, with $\hat{\Omega}_{W,i} = \text{blockdiag}(0, \hat{\Omega}_k)$ and $\hat{W}_i^\circ = (\hat{D}_i^\circ, \hat{S}_i^{\circ\prime})'$, we compute

$$\hat{A} = N^{-1}\sum_i \hat{W}_i^\circ \hat{W}_i^{\circ\prime} - N^{-1}\sum_i \hat{\Omega}_{W,i}, \qquad \hat{b} = N^{-1}\sum_i \hat{W}_i^\circ \hat{Y}_i^\circ, \qquad (12)$$

$$\hat{a} = \hat{A}^{-1}\hat{b} = (\hat{\theta}_C, \hat{\beta}_C')'. \qquad (13)$$

Equation (13) is the Croon-DML estimator; because equation (10) has no intercept, equation (12) uses uncentered second moments. We fix the factor variances to 1 so that $\hat{\beta}_C$, $\hat{\gamma}$, and the residualized reliabilities are comparable across factors ($\theta_0$ is invariant to the scaling of the latent variables), and fix each factor's direction by the sign of its reference loading. The measurement model is refitted in each training fold so that no held-out observation enters its own score weights; the fold-specific fits differ from a full-sample fit by $O_p(N^{-1/2})$. A nonpositive-definite corrected latent block or $\hat{A}$ is recorded as an estimation failure (Section S4.2).

**Table 2.** Cross-fitted estimation procedure for Croon-DML

| Step | Data | Computation | Output |
|---|---|---|---|
| (1) Measurement model | Training folds $I_k^c$ | Fit the prespecified measurement model by normal-theory ML and compute equations (4) and (6) | $\hat{\tau}_k, \hat{\Lambda}_k, \hat{\Theta}_k; \hat{B}_k, \hat{\Omega}_k$ |
| (2) Factor scores | Training data $I_k^c$ and held-out data $I_k$ | Compute $\hat{S}_{ik} = \hat{B}_k(X_i - \hat{\tau}_k)$ for both data sets with the parameters from step (1) | Training and held-out factor scores |
| (3) Conditional means and residuals | Learned in the training folds, applied to the held-out fold | Regress $Y, D, S$ on $Z$ (learner) to obtain $\hat{\ell}_k, \hat{m}_k, \hat{q}_k$, and compute residuals in the held-out fold | $\hat{Y}_i^\circ, \hat{D}_i^\circ, \hat{S}_i^\circ$ |
| (4) Second-moment correction | All held-out folds pooled | Equation (12): subtract $\hat{\Omega}_W$ from the second moments of the residuals (Croon-DML-R uses $\hat{\Omega}_{W,i}^R$ of equation (15)) | $\hat{A}, \hat{b}$ |
| (5) Estimation | — | Equation (13); record an estimation failure if $\hat{A}$ is not positive definite | $\hat{a} = (\hat{\theta}_C, \hat{\beta}_C')'$ |

*Note.* Steps (1)–(3) are repeated for $k = 1, \dots, K$, with scale identification and factor direction fixed by the same rule across folds. Held-out observations are not used to estimate $\hat{B}_k, \hat{\Omega}_k$ or to learn the conditional means. ML = maximum likelihood.

### *3.4 Orthogonality and consistency*

**Theorem 2 (Neyman orthogonality with respect to the conditional means).** At the true structural parameters, the expectation of equation (10) is unchanged to first order by small perturbations of $\ell_0, m_0, q_0$: under A2 and the conditions on the structural errors, $E(W^\circ \mid Z) = 0$ and $E(e - U'\beta_0 \mid Z) = 0$, so all three derivatives are zero (Section S1.3). It is not orthogonal with respect to $\Omega$: the derivative in direction $\Delta\Omega$ is $\text{blockdiag}(0, \Delta\Omega)a_0$.

**Theorem 3 (Consistency of the point estimator).** Suppose that A1–A3 and equations (1)–(2) hold for independent and identically distributed observations with $J$ and $K$ fixed. If the measurement model estimator in each training fold is consistent under the fixed scale and direction, the cross-fitted conditional-mean estimators converge to the true conditional means in mean square, and $E(Y^2 + D^2 + \| X \|^2) < \infty$, then $\hat{a} \rightarrow_p a_0$ (Section S1.4).

**Theorem 4 (Asymptotic normality).** Suppose that the conditions of Theorem 3 hold, that the products of the $L_2$ learning errors of $\hat{\ell}_k, \hat{m}_k, \hat{q}_k$ and $\| \hat{q}_k - q_0 \|_2^2$ are $o_p(N^{-1/2})$, that the measurement-parameter estimator of each training fold is asymptotically linear with influence function $\psi(X)$, and that fourth moments are finite. Then $\sqrt{N}(\hat{a} - a_0) \rightarrow_d N(0, A_0^{-1} V A_0^{-1})$ with $V = \mathrm{Var}\{g_i(a_0) + J_\vartheta \psi(X_i)\}$, where $J_\vartheta$ is the derivative of $E\{g(a_0)\}$ with respect to the measurement parameters (Section S1.4): the conditional means contribute no first-order term (Theorem 2), the measurement parameters contribute through $J_\vartheta$. Its sample analogue is the *stacked sandwich* SE (Stefanski & Boos, 2002); the *conditional sandwich* SE omits $J_\vartheta \psi$ (Section 5.5).

***3.5 Residual bias after correction: Signal learning error and noise fitting***

Two sources of bias remain after the $\Omega$ correction, both from learning the conditional means of the factor scores: learning error in the latent signal, and *noise fitting*, in which the learner also fits the training-fold score errors, so that the error variance of the residualized scores exceeds $\Omega$ even under cross-fitting. Croon-DML subtracts only $\Omega$, so both leave residual bias in the direction of the uncorrected bias. We analyze them for a *linear smoother* (Buja et al., 1989), a learner whose prediction is a weighted sum $w(Z)' t_{\text{train}}$ of the training target values; when the weights do not depend on the target values and the same weights serve all three conditional means, we call it a *common prespecified linear smoother*. The *inflation factor* $1 + \| w(Z_i) \|^2$ multiplies the second moment of the noise in the residual of held-out observation $i$ (Hastie et al., 2009), and $\bar{\omega} = E\{\| w(Z) \|^2\}$ is the mean sum of squared weights.

**Proposition 1 (Bias of the population solution due to learning error in $q_0$).** Suppose that the measurement model and $\ell_0, m_0$ are known and only $q_0$ is estimated from an independent training sample, and let $\delta_q(Z) = \hat{q}(Z) - q_0(Z)$ and $\Delta_q = E\{\delta_q(Z)\delta_q(Z)' \mid \text{training sample}\}$. Then the latent block of the corrected second moment is $\Sigma_\eta + \Delta_q$, and the solution of this conditional population estimating equation differs from $\theta_0$ by equation (9) with $\Omega$ replaced by $\Delta_q$ (Section S1.5); if $\| \hat{q} - q_0 \|_2 = O_p(r_N)$ and the denominator is bounded away from zero, this bias is $O_p(r_N^2)$.

Because $\hat{q}$ regresses $S = \eta + U$ on $Z$, $\delta_q = \delta_\eta + \delta_U$ splits into the learning error in the latent signal and the noise fit, and $\Delta_q = \Delta_\eta + \Delta_U$ (Section S1.6). $\Delta_\eta$ arises in the same form for the oracle and largely cancels under a common linear smoother: applying the residualization operator $R$ ($Rt_i = t_i - w(Z_i)' t_{\text{train}}$) to the structural model gives

$$RY = \theta_0\, RD + \beta_0' RS - \beta_0' RU + Rh_{Y0} + Re, \qquad (16)$$

so the signal learning error in $q_0$ enters $RY$ through the same coefficient $\beta_0$ and cancels in the estimation of $\theta_0$; only the noise term $-\beta_0' RU$ and second-order products of unremoved structural-function signals, shared by the oracle, remain. With target-specific smoothers $R_Y, R_D, R_S$, the terms $\theta_0(R_Y - R_D)D + \beta_0'(R_Y - R_S)\eta$ do not cancel (smoother mismatch). $\Delta_U$ arises only for the factor scores and grows with the flexibility of the learner: for a linear smoother, the residualized score of held-out row $i$ carries the noise $U_i - w(Z_i)' U_{\text{train}}$, with covariance $\Omega\{1+\| w(Z_i) \|^2\}$. Under three conditions, its magnitude has a closed-form approximation.

- C1. A common prespecified linear smoother $w$ is used for all three conditional means.
- C2. The measurement parameters are known, and the conditional second moments of $(\eta^\circ, v, e, U)$ do not vary with $Z$ (not needed for identification or consistency).
- C3. The structural-function signal not removed by each conditional-mean fit is negligible.

**Proposition 2 (Approximate residual bias of Croon-DML).** Under C1–C3, the difference between the population solution of Croon-DML and $\theta_0$ is approximated by equation (9) with $\Omega$ replaced by

$$\Delta_U^* = \frac{\bar{\omega}}{1+\bar{\omega}}\Omega, \qquad \bar{\omega} = E\{\| w(Z) \|^2\} \qquad (14)$$

(Section S1.6). $\bar{\omega}$ grows as the penalty weakens or the number of terms approaches the sample size.

The same derivation modifies the correction term: we replace $\widehat{\Omega}_{W,i}$ in equation (12) with

$$\widehat{\Omega}_{W,i}^R = \text{blockdiag}\{0,\ \widehat{\Omega}_k \circ (1 + G_i)\}, \qquad (G_i)_{jk} = w_j(Z_i)'w_k(Z_i), \qquad (15)$$

where $\circ$ denotes the Hadamard product, 1 the $J \times J$ matrix of ones, and $w_j$ the smoother weights for the $j$th factor-score component (Section S4.1). We call the resulting estimator the residualization-inflation-corrected Croon-DML (Croon-DML-R).

**Proposition 3 (Population solution of Croon-DML-R).** Under C1–C3, the noise in $D^\circ, Y^\circ$ is inflated by the same factor, so the population solution of Croon-DML-R is $a_0$ (Section S1.6). Equation (15) removes the noise component exactly even when the penalty differs across factor-score components; the smoother mismatch in the signal remains and is addressed through the choice of learner. With penalties selected by cross-validation, equations (14) and (15) treat the selected penalties as fixed.

## 4. Simulation design

### *4.1 Data generation*

We generate observed covariates $Z \sim N(0, I_{20})$ and three latent covariates $\eta = q_0(Z) + \zeta = \Pi Z + \zeta$. $\Pi$ links $Z_1$–$Z_6$, two at a time, to the three latent variables with coefficients 0.25, so $q_0$ is linear and the observed covariates explain 12.5% of each latent variance (Figure S1 varies this proportion); $\text{Var}(\zeta) = \Sigma_\eta$ gives each latent variable unconditional variance 1 and latent correlations of .30. The focal predictor and the outcome are

$$D = f_D(Z) + \eta'\gamma_0 + v, \qquad Y = \theta_0 D + f_Y(Z) + \eta'\beta_0 + e, \qquad (16)$$

$$f_D(Z) = 0.50Z_1 + 0.35Z_2^2 + 0.40Z_3^2 + 0.30Z_4Z_5, \quad f_Y(Z) = 0.50Z_1 - 0.30Z_2^2 + 0.35Z_6^2 + 0.25Z_7Z_8, \qquad (17)$$

with $v, e \sim N(0,1)$, $\theta_0 = 0.30$, and $\beta_0 = (0.40, 0.30, 0.20)'$.

**Table 3.** Main Monte Carlo design

| Design factor | Setting |
|---|---|
| Sample size | $N = 500, 1000, 5000$ |
| Measurement quality | Lower, higher (Table 4) |
| Confounding strength | Moderate, strong: $\gamma_0 = (0.25, 0.20, 0.15)'$, $(0.50, 0.40, 0.30)'$ |
| Conditions and replications | 12 conditions, 1,000 replications per condition |

| Design factor | Setting |
|---|---|
| Observed and latent covariates | 20 observed covariates; three latent covariates (four indicators each) |
| True coefficient of interest | $\theta_0 = 0.30$ |
| Conditional-mean learning | Quadratic polynomial ridge regression with five-fold cross-fitting; penalty by target-specific LOOCV or by common-penalty LOOCV (recommended) |
| Estimators compared | Oracle latent-variable DML, uncorrected factor-score DML, Croon-DML, Croon-DML-R |

*Note.* LOOCV = leave-one-out cross-validation; DML = double/debiased machine learning.

### *4.2 Measurement conditions and estimators compared*

Each latent variable has four continuous unit-variance indicators (error variances $1 - \lambda^2$); Table 4 gives the loadings, factor-score error variances, reliabilities, and equation (9) bias for both levels of measurement quality. The fitted measurement model fixes the factor variances to 1 and means to 0, estimates the factor correlations, and includes no indicator error covariances; the four estimators (Table 3) use the same data, folds, measurement model, and learner.

**Table 4.** Standardized loadings, factor-score error variances, reliabilities, and uncorrected bias by measurement condition

| Measurement quality | Standardized loadings (factor 1 / 2 / 3) | $\mathrm{diag}(\Omega)$ | Bartlett-score reliability before residualization | Residualized reliability | Bias from equation (9): moderate / strong confounding |
|---|---|---|---|---|---|
| Lower | .70, .66, .68, .64 / .55, .52, .58, .50 / .42, .45, .40, .38 | (0.304, 0.608, 1.209) | (.77, .62, .45) | (.74, .59, .42) | 0.067 / 0.119 |
| Higher | .88, .85, .86, .83 / .75, .72, .76, .70 / .62, .65, .60, .58 | (0.090, 0.213, 0.412) | (.92, .82, .71) | (.91, .80, .68) | 0.028 / 0.054 |

*Note.* $\Omega$ = Bartlett factor-score error covariance. Bartlett-score reliability before residualization = $1/(1 + \mathrm{diag}\Omega)$ (factor variances equal 1); residualized reliability = $\mathrm{diag}(\Sigma_\eta)/\mathrm{diag}(\Sigma_\eta + \Omega)$ (Section 2.4). Bias from equation (9) = analytic bias of uncorrected factor-score DML (DML = double/debiased machine learning).

### *4.3 Learners and performance measures*

The learner is ridge regression (Hoerl & Kennard, 1970) on the 230 terms of the second-degree polynomial expansion of the observed covariates; the functions in equation (17) lie in this model space, so the conditional means are not misspecified. The penalty parameter $\alpha$ is selected in each fold by leave-one-out cross-validation (LOOCV) from 16 log-spaced candidates, either separately for each target (target-specific LOOCV) or jointly for the three conditional means (common-penalty LOOCV, the recommended learner: one $\alpha$ minimizing the sum of the standardized target-specific LOOCV errors). Two further settings serve the diagnostics: ridge regression with $\alpha = 1$ fixed (satisfying C1) and, in the design with 100 observed covariates, gradient boosting and random forests (Section S4.1). The noise-fit ratio $\mathrm{tr}(\widehat{\Delta}_U)/\mathrm{tr}(\Omega)$ is computed by applying the selected weights to the training factor-score errors; in applications, where $U$ is unobserved, the design-based ratio $\mathrm{tr}(\bar{G} \circ \hat{\Omega})/\mathrm{tr}(\hat{\Omega})$, with $\bar{G}$ the mean of the $G_i$, replaces it. Performance is evaluated by bias, empirical standard deviation (SD), root mean square error (RMSE), and Monte Carlo standard error (MCSE), following Morris et al. (2019). Nonconvergence and improper solutions are not masked by constraints or retries but recorded as estimation failures (F. Chen et al., 2001).

### *4.4 Supplementary analyses*

The supplementary analyses (Tables S1–S9 and S13–S19) decompose the residual bias, compare the penalty rules in paired runs, examine violations of A2, misspecification, a nonlinear $q_0$, and 100 observed covariates, and evaluate Wald intervals from the conditional and stacked sandwich standard errors (SEs) and percentile and bias-corrected (BC) intervals from a 300-resample bootstrap that refits the measurement model (Efron, 1987; Efron & Tibshirani, 1993). Data, materials, and code are available in Harvard Dataverse (https://doi.org/10.7910/DVN/E2EWGC).

## 5. Results

### *5.1 Performance of the factor-score error correction*

Table 5 reports bias and RMSE for the 12 conditions under both penalty rules, and Figure 2 shows them for the recommended learner (no estimation failures). The bias of uncorrected DML ranged from 0.028 to 0.118, close to the values from equation (9), and did not decrease with the sample size. The bias of Croon-DML ranged from 0.001 to 0.022 and the absolute bias of the oracle from 0.000 to 0.014; the oracle's own bias is the second-order learning error of $\ell_0$ and $m_0$ that any partially linear DML carries at these sample sizes. In all 12 conditions, Croon-DML reduced the absolute bias of uncorrected DML by 66–95% and its RMSE by 12–84% (computed from unrounded estimates) despite a slightly larger SD. With the same target-specific learner, Croon-DML-R changed little (bias 0.001–0.020, within 0.004 of Croon-DML), because noise fitting is small under the large penalties selected by target-specific LOOCV (Section 5.3).

The last three columns of each panel use the recommended learner with one common penalty for the three conditional means. The maximum gap between Croon-DML and the oracle fell from 0.014 (target-specific penalties) to 0.011, and Croon-DML-R reduced it to at most 0.004 ($N \leq 1000$, lower quality, strong confounding) and to 0.000 at $N = 5000$; the common penalty also reduced the RMSE of these three estimators by 0.003–0.006 at $N \leq 1000$. The common penalty lowers the penalty of the factor-score targets (from about 2,100–3,200 to about 850–1,000), which reduces signal learning error but increases noise fitting, and Croon-DML-R removes the latter; the two devices close the gap only together.

**Table 5.** Bias and root mean square error in the main Monte Carlo study ($\theta_0 = 0.30$, 1,000 replications per condition)

**A. Bias**

| $N$ | Measurement quality | Confounding | Oracle | Uncorrected | Croon-DML | Croon-DML-R | Oracle (common) | Croon-DML (common) | Croon-DML-R (common) |
|---|---|---|---|---|---|---|---|---|---|
| 500 | Lower | Moderate | 0.008 | 0.063 | 0.014 | 0.013 | 0.010 | 0.014 | 0.011 |
| 500 | Lower | Strong | 0.011 | 0.106 | 0.022 | 0.020 | 0.009 | 0.018 | 0.012 |

| $N$ | Measurement quality | Confounding | Oracle | Uncorrected | Croon-DML | Croon-DML-R | Oracle (common) | Croon-DML (common) | Croon-DML-R (common) |
|---|---|---|---|---|---|---|---|---|---|
| 500 | Higher | Moderate | 0.008 | 0.031 | 0.011 | 0.010 | 0.009 | 0.011 | 0.010 |
| 500 | Higher | Strong | 0.014 | 0.055 | 0.018 | 0.017 | 0.011 | 0.014 | 0.012 |
| 1000 | Lower | Moderate | 0.008 | 0.069 | 0.014 | 0.012 | 0.009 | 0.013 | 0.009 |
| 1000 | Lower | Strong | 0.008 | 0.113 | 0.022 | 0.020 | 0.006 | 0.017 | 0.010 |
| 1000 | Higher | Moderate | 0.007 | 0.032 | 0.010 | 0.009 | 0.008 | 0.010 | 0.009 |
| 1000 | Higher | Strong | 0.009 | 0.056 | 0.014 | 0.013 | 0.007 | 0.010 | 0.007 |
| 5000 | Lower | Moderate | 0.001 | 0.068 | 0.004 | 0.002 | 0.001 | 0.005 | 0.002 |
| 5000 | Lower | Strong | 0.001 | 0.118 | 0.008 | 0.004 | 0.001 | 0.008 | 0.001 |
| 5000 | Higher | Moderate | 0.000 | 0.028 | 0.001 | 0.001 | 0.000 | 0.002 | 0.001 |
| 5000 | Higher | Strong | 0.002 | 0.054 | 0.004 | 0.002 | 0.001 | 0.004 | 0.002 |

**B. Root mean square error**

| $N$ | Measurement quality | Confounding | Oracle | Uncorrected | Croon-DML | Croon-DML-R | Oracle (common) | Croon-DML (common) | Croon-DML-R (common) |
|---|---|---|---|---|---|---|---|---|---|
| 500 | Lower | Moderate | 0.049 | 0.080 | 0.053 | 0.053 | 0.045 | 0.049 | 0.049 |
| 500 | Lower | Strong | 0.047 | 0.115 | 0.057 | 0.057 | 0.043 | 0.052 | 0.051 |
| 500 | Higher | Moderate | 0.050 | 0.059 | 0.052 | 0.051 | 0.046 | 0.047 | 0.047 |
| 500 | Higher | Strong | 0.050 | 0.072 | 0.053 | 0.052 | 0.046 | 0.049 | 0.048 |
| 1000 | Lower | Moderate | 0.035 | 0.077 | 0.040 | 0.040 | 0.032 | 0.036 | 0.035 |
| 1000 | Lower | Strong | 0.034 | 0.118 | 0.044 | 0.043 | 0.031 | 0.039 | 0.037 |
| 1000 | Higher | Moderate | 0.035 | 0.048 | 0.037 | 0.037 | 0.032 | 0.033 | 0.032 |
| 1000 | Higher | Strong | 0.034 | 0.065 | 0.038 | 0.037 | 0.031 | 0.033 | 0.033 |
| 5000 | Lower | Moderate | 0.015 | 0.069 | 0.017 | 0.017 | 0.015 | 0.017 | 0.016 |
| 5000 | Lower | Strong | 0.014 | 0.119 | 0.019 | 0.018 | 0.014 | 0.019 | 0.017 |
| 5000 | Higher | Moderate | 0.015 | 0.032 | 0.016 | 0.016 | 0.015 | 0.016 | 0.016 |
| 5000 | Higher | Strong | 0.014 | 0.056 | 0.016 | 0.016 | 0.014 | 0.016 | 0.015 |

*Note.* Bias = mean estimate − $\theta_0$; RMSE = root mean square error; oracle = oracle latent-variable DML; uncorrected = uncorrected factor-score DML; Croon-DML-R = residualization-inflation correction (equation 15). Columns without "(common)" use quadratic polynomial ridge regression with target-specific LOOCV; columns with "(common)" use the same learner with one common penalty for the three conditional means (the recommended learner; the uncorrected estimator under the common penalty, bias 0.028–0.116, is given in Table S15). All estimators use the same data, folds, and measurement model. Monte Carlo standard errors of bias were all 0.002 or smaller, and the estimation failure rate was .000. The analytic bias from equation (9) is given in Table 4. DML = double/debiased machine learning; LOOCV = leave-one-out cross-validation.

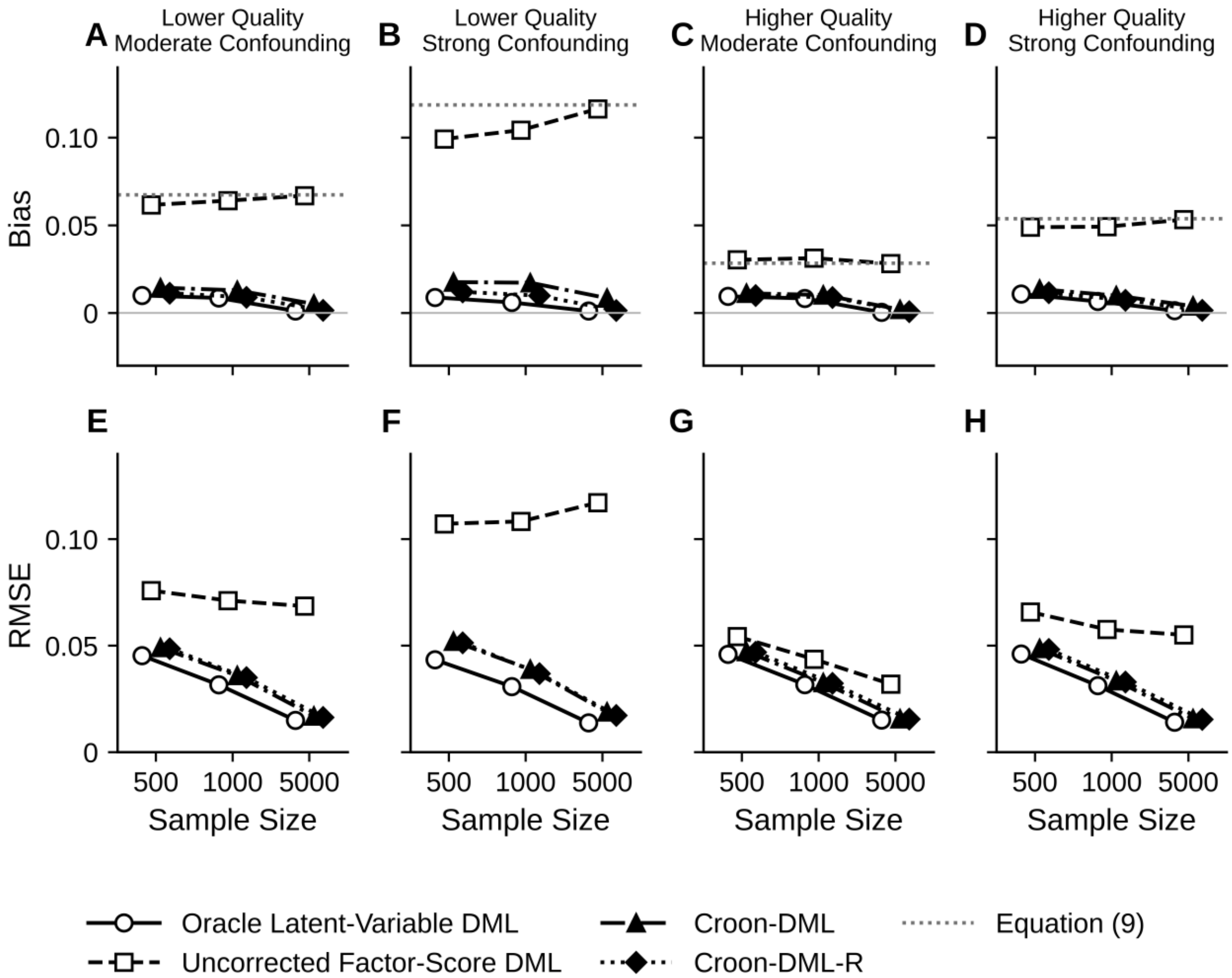

**Figure 2.** Bias (A–D) and RMSE (E–H) of the four estimators by sample size in the main Monte Carlo study (common-penalty LOOCV, the recommended learner). Panel columns are the combinations of measurement quality (lower/higher) and confounding strength (moderate/strong). Circles denote oracle latent-variable DML, squares uncorrected factor-score DML, triangles Croon-DML, and diamonds Croon-DML-R; the dotted line in the bias panels is the analytic uncorrected bias from equation (9). Monte Carlo standard errors (MCSE) of bias are at most 0.002, smaller than the plotting symbols. DML = double/debiased machine learning; LOOCV = leave-one-out cross-validation; RMSE = root mean square error.

***5.2 Sources of residual bias***

The bias that remains in Croon-DML stems from learning the conditional means: it was close to zero with the true conditional means but remained almost entirely with the true measurement parameters and learned conditional means (Table S1), and the bias with only $q_0$ learned was close to the population solution of Proposition 1 and larger than with all three conditional means learned (Table S2; Section 3.5). Target-specific LOOCV selected large penalties for the factor-score targets (about 2,100–3,200), which carry little signal, and smaller penalties for $D$ and $Y$; the noise-fit ratio was only 0.013–0.025 (Tables S2 and S6), so the residual bias stems from signal learning error and smoother mismatch rather than noise fitting.

Because the selected penalty stays in this range regardless of $N$ while the eigenvalues of the training design grow with $N$, the learned conditional means retained only about 10%, 20%, and 60% of $q_0(Z)$ (the slope of the learned conditional mean on $q_0(Z)$) at $N = 500{,}1000{,}5000$ with lower quality (25%, 40%, and 80% under the common penalty); Proposition 1 captures this regularization bias as $\Delta_q$. It falls at about the rate of the SD (bias/SD of Croon-DML with lower quality and strong confounding: 0.41, 0.57, 0.46 at $N = 500{,}1000{,}5000$; oracle 0.24, 0.24, 0.09), so the selected penalties violate the rate condition of Theorem 4 at these sample sizes and Wald intervals fall short of nominal coverage even with accurate SEs (Section 5.5).

### *5.3 Noise fitting and Croon-DML-R*

On the same data and folds, ridge regression with $\alpha = 1$ fixed produced substantial noise fitting (Tables S6 and S14; Figure 3): with lower quality and strong confounding, the bias of Croon-DML was 0.090, 0.049, and 0.012 at $N = 500{,}1000{,}5000$, against 0.023, 0.021, and 0.008 under LOOCV. The approximation from equation (14) matched the observed Croon-DML − oracle gap within 0.007 in all 12 conditions (Figure 3A), and the gap grew with the noise-fit ratio (Figure 3B). Croon-DML-R (equation 15) eliminated it, bringing the bias within 0.003 of the oracle's in all 12 conditions and, with lower quality, strong confounding, and $N = 1000$, the RMSE from 0.066 to 0.050 (Table S14).

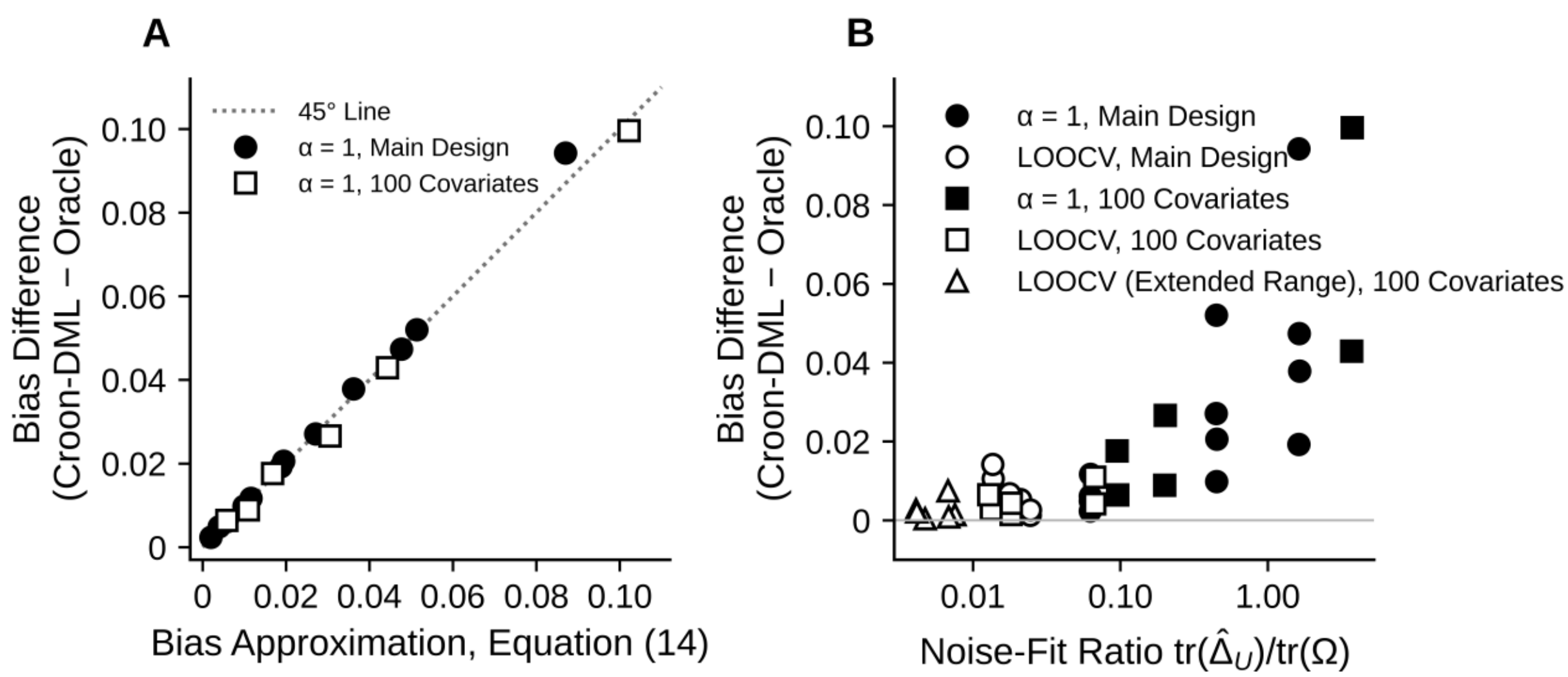

**Figure 3.** Noise fitting and residual bias after correction. (A) Bias difference Croon-DML − oracle (vertical axis) versus the bias approximation based on equation (14) (horizontal axis) under a common prespecified smoother ($\alpha = 1$). Filled circles are the 12 conditions of the main design and open squares the six conditions with 100 observed covariates; the dotted line is the 45° line. (B) The same gap as a function of the noise-fit ratio $\mathrm{tr}(\widehat{\Delta}_U)/\mathrm{tr}(\Omega)$ (log scale) in the ridge regression conditions for which noise fitting was diagnosed. Filled symbols denote $\alpha = 1$ and open symbols LOOCV (leave-one-out cross-validation; triangles, extended candidate range).

### *5.4 Robustness: Assumption violations, misspecification, and nonlinear and high-dimensional conditions*

Under violations of A2, the bias of Croon-DML shifted in the negative direction predicted by equation (S11) as $E(Uv)$ increased, and under strong violations its absolute bias exceeded that of uncorrected DML (Table S3). Linear ridge regression for $m_0$, the common practice of adjusting for observed covariates linearly, biased Croon-DML and the oracle alike ($-0.12$ to $-0.13$; Table S4). Omitting an indicator error covariance (error correlation .30) from the measurement model increased the bias from 0.009 to 0.027 even at $N = 5000$ (Table S5), and omitting the quadratic terms of a nonlinear $q_0$ left a bias of 0.038–0.039 at $N = 5000$ for both the oracle and Croon-DML (Table S13).

With 100 observed covariates and nonlinear conditional means, Croon-DML had smaller absolute bias than uncorrected DML for every learner (Table S7); with $\alpha = 1$ applied to 5,150 second-degree terms, which exceed the training sample size at $N = 5000$, the noise-fit ratio reached 3.7 and the bias of Croon-DML (0.104) approached the uncorrected bias (0.120), as equation (14) predicts (Figure 3A).

### *5.5 Interval estimation*

For the recommended combination—common-penalty LOOCV, Croon-DML-R, and a bootstrap that refits the measurement model with the penalties fixed at the original-sample values (Section S3.2)—95% Wald intervals from the conditional and from the stacked sandwich SE both covered $\theta_0$ in .934–.962 of the samples and BC bootstrap intervals in .930–.962, with Type I error rates at $\theta_0 = 0$ of .038–.064, .038–.062, and .038–.062 (Tables S18–S19). The measurement parameters contributed at most 2% of the stacked variance, so the stacked SE, which needs no bootstrap, suffices. With target-specific penalties, the conditional sandwich SE of Croon-DML was accurate (SE/SD 0.91–1.01), but its Wald coverage was only .872–.936 because of the residual bias (Section 5.2), and that of uncorrected DML was zero at $N = 5000$ (Tables S8–S9). Bootstrap SEs exceeded the SD by 4–29% for all estimators because duplicated resample observations in the same fold reduce the effective training sample size (Section S3.2).

## 6. Empirical illustration

Using public data from the High School Longitudinal Study of 2009 (HSLS:09; National Center for Education Statistics, 2018), we illustrate the four steps recommended in Section 7 and the diagnosis of the size of the correction. The analysis describes a conditional association in an unweighted complete-case sample ($N = 9{,}491$ of 23,503 students) and does not test an effect. The outcome $Y$ is the 2012 mathematics identity scale score, the focal predictor $D$ weekly time spent on mathematics homework and study in 2012 (representative hours, 0–11), the latent covariates 2009 mathematics identity (two items) and mathematics self-efficacy (four items), and the observed covariates 33 variables from 2009, including prior mathematics achievement (X1TXMTH; Table S10).

The correlated two-factor measurement model fit well: $\chi^2(8) = 259.8$ was significant because of the large sample, but the Tucker–Lewis index = .986, comparative fit index = .993, root mean square error of approximation = .058 (inflated at small degrees of freedom; Kenny et al., 2015), and standardized root mean square residual = .014 indicated good fit (Hu & Bentler, 1999). Standardized loadings were .79–.90, the composite reliabilities (Raykov, 1997) .845 and .897, and the average variance extracted (.73, .69) exceeded the squared factor correlation (.44; Fornell & Larcker, 1981). Configural, metric, and scalar invariance across gender held (Hong et al., 2003; F. F. Chen, 2007), and a moderated nonlinear factor analysis with prior achievement as the moderator (Bauer, 2017), which examines the first condition of A2, left the correction unchanged (Section S5, Tables S11–S12). The diagonal elements of $\widehat{\Omega}$, averaged over training folds, were 0.164 and 0.110, and the residualized reliabilities .83 and .89. The conditional means were learned with the recommended learner (Section S8.1); target-specific penalties gave the same estimates to within 0.0001.

Table 6 compares the four estimators under the same residualization. With all 33 observed covariates, the latent-covariate adjustment was 0.0071 and the factor-score error correction 0.0010, 15% of the adjustment, with a bias-corrected interval whose lower limit was zero. Evaluating equation (9) at the sample moments (Section S1.1) attributes the small correction not to the reliabilities (.83, .89) but to $\hat{\gamma} = (0.057, 0.008)'$ (with $\hat{\beta}_C = (0.56, 0.02)'$): once prior mathematics achievement is controlled, the latent covariates are nearly unrelated to study time. To represent a study without a strong observed proxy, we excluded prior achievement; $\hat{\gamma}$ then rose to $(0.128, 0.016)'$, the adjustment to 0.0194, and the correction to 0.0026 (13%). Croon-DML-R agreed with Croon-DML to within 0.0001 because noise fitting was small (mean inflation factor 1.007); the stacked and conditional sandwich SEs were both 0.0043 and the bootstrap SE 0.0045. In standardized terms, the Croon-DML-R coefficient is .041 SD (with prior achievement) or .054 SD (without) of the identity scale score per 1 SD (2.07 hours) of study time, a small association.

The 2009 identity indicators are the same items as the 2012 outcome (a repeated measurement), so stable item-specific components would violate A2 ($E(Ue) \neq 0$; equation (S11)); with 2012 mathematics achievement, which shares no items with the indicators, as the outcome, the correction remained 15% and 13% of the adjustment (Table S17). Because complete cases are 40% of the sample, a multiple-imputation sensitivity analysis (17,790 students with observed 2012 outcome and study time; Rubin, 1987; 10 imputations) was added; its corrections were 0.0014 and 0.0028 (Table S16). The illustration has the usual limitations of secondary data (measurement error in prior achievement and study time; four-point items treated as continuous, Rhemtulla et al., 2012; Lu & Thomas, 2008; survey weights and school clustering ignored), and its magnitudes are specific to this 2009 U.S. cohort.

**Table 6.** Adjusted association between mathematics study time and 2012 mathematics identity in HSLS:09, by estimator and observed-covariate specification

| Estimator | With prior achievement (33) | Without prior achievement (32) |
|---|---|---|
| DML without latent covariates | 0.0275 (0.0049) [0.0185, 0.0374] | 0.0462 (0.0050) [0.0363, 0.0558] |
| Uncorrected factor-score DML | 0.0215 (0.0044) [0.0133, 0.0306] | 0.0294 (0.0044) [0.0217, 0.0386] |
| Croon-DML | 0.0205 (0.0045) [0.0120, 0.0299] | 0.0268 (0.0046) [0.0184, 0.0362] |
| Croon-DML-R | 0.0205 (0.0045) [0.0120, 0.0299] | 0.0268 (0.0046) [0.0184, 0.0361] |
| Latent-covariate adjustment (no latent − Croon-DML-R) | 0.0071 [0.0018, 0.0123] | 0.0194 [0.0129, 0.0250] |
| Factor-score error correction (uncorrected − Croon-DML-R) | 0.0010 [0.0000, 0.0022] | 0.0026 [0.0015, 0.0039] |

*Note.* $N = 9{,}491$. Cells show the point estimate (bootstrap SE) [bias-corrected 95% interval]; for the adjustment and the correction, the interval is the bias-corrected 95% interval of the difference between the two estimates computed in the same resamples. The bootstrap used 300 resamples (no failures); the penalty and fold rules are described in Section S8.1. The conditional means were learned by quadratic polynomial ridge regression with one common LOOCV penalty for the three conditional means (five-fold cross-fitting); the mean inflation factor of Croon-DML-R was 1.007 (with prior achievement) and 1.006 (without). All values are conditional associations

in the unweighted complete-case sample. DML = double/debiased machine learning; HSLS:09 = High School Longitudinal Study of 2009; LOOCV = leave-one-out cross-validation; SE = standard error.

**7. Discussion**

Factor-score error and conditional-mean learning are distinct sources of error, and the remedies are not interchangeable. The $\Omega$ correction removes the error originating in the measurement model and accounts for most of the improvement, and the common penalty removes most of the remaining signal learning error at the price of more noise fitting. Croon-DML-R removes the noise-fitting term of a linear smoother (exactly when its weights are fixed; Proposition 3); this matters when the penalty is weak or the number of terms approaches the sample size, and costs nothing otherwise (its SD equals that of Croon-DML to within 0.001). Nor do alternatives remove the need for the correction. A full structural equation model requires the functions of the observed covariates to be specified parametrically. Posterior-mean factor scores (Nguyen & Stuart, 2020) avoid the bias only under a correctly specified parametric model for $\eta$ given $D$, $Y$, and $Z$, and their error, correlated with the latent variable, is not removed by subtracting $\Omega$. Indicators entered as observed covariates keep their measurement error.

For applied research, we recommend four steps. First, specify a theory-based measurement model and examine its fit, composite reliability, average variance extracted, and measurement invariance (Anderson & Gerbing, 1988). Second, compute $\widehat{\Omega}$ and the residualized reliabilities, but do not omit the correction when reliability is high, because its size also depends on the latent-covariate coefficients. Third, report, with the same learner, the estimates without latent covariates, with uncorrected factor scores, and with the correction, and decompose the correction with equation (9). Fourth, with a linear smoother such as ridge regression, choose the penalty by prediction error but use one common penalty for the three conditional means (target-specific tuning creates smoother mismatch), remove noise fitting with Croon-DML-R, and obtain intervals from the stacked sandwich SE or, at greater cost, from a BC bootstrap that also refits the measurement model.

Three limitations concern the method. The stacked sandwich SE treats the smoother weights and the selected penalties as fixed, and the asymptotic theory assumes the rate conditions of Theorem 4. Croon-DML-R is defined only for linear smoothers, and the common-penalty rule was examined only in the main design; with tree learners, which provide no weights, noise fitting can only be limited by regularization (Table S7). The latent covariates enter linearly; latent nonlinear effects (cf. Rosseel et al., 2025) and combinations with full-information maximum likelihood or multiple imputation remain open. A corrected latent block that is not positive definite (seen only in bootstrap resamples at $N = 500$ with lower measurement quality) calls for better measurement or more data, not a constrained fit.

**8. Conclusion**

Subtracting the error covariance of Bartlett factor scores from the cross-fitted estimating equation removes most of the bias of partially linear DML with latent covariates. A common penalty and Croon-DML-R remove most of the bias that learning the conditional means of the factor scores leaves behind, and the stacked sandwich SE provides inference without a bootstrap.

**Data and Code Availability Statement**

Data, materials, and code are available in Harvard Dataverse (https://doi.org/10.7910/DVN/E2EWGC). The empirical example uses the public High School Longitudinal Study of 2009 data (National Center for Education Statistics, 2018).

**Disclosure of the Use of Artificial Intelligence-Generated Content (AIGC) Tools**

During preparation of this work, the authors used DeepL to translate the initial draft and improve English readability, and OpenAI Codex to assist with refinement and error checking of the research code. The authors reviewed and edited all outputs and take full responsibility for the final manuscript and research code.

**Acknowledgments**

None. No funding was received for this study.

**Disclosure of Interest**

The authors declare no competing interests.

**Ethics Statement**

This study used only computer-simulated data and publicly available, de-identified secondary data (the HSLS:09 public-use file). The study involved no intervention or interaction with individuals, and the authors had no access to information that could reasonably be used to identify an individual directly or indirectly. The authors did not link the data to external individual-level records or attempt to identify any individual. The study therefore did not constitute human subjects research under the Bioethics and Safety Act of the Republic of Korea. Ethical approval was not required.

## Supporting Information

### S1. Theoretical derivations

***S1.1 Bias of the uncorrected structural coefficient (equation (9))***

From equations (1), (2), and (7),

$$E(S^\circ S^{\circ\prime}) = \Sigma_\eta + \Omega,\quad E(S^\circ D^\circ) = \Sigma_\eta \gamma_0,\quad E(S^\circ Y^\circ) = \Sigma_\eta(\theta_0\gamma_0 + \beta_0),\quad E(D^{\circ 2}) = C_{DD},\quad E(D^\circ Y^\circ) = \theta_0 C_{DD} + \gamma_0' \Sigma_\eta \beta_0. \qquad (1)$$

The uncorrected estimator is the first component of $[E(W^\circ W^{\circ\prime})]^{-1}E(W^\circ Y^\circ)$; the inverse of a partitioned matrix gives

$$\theta_{\mathrm{U}} = \frac{E(D^\circ Y^\circ) - E(D^\circ S^{\circ\prime})E(S^\circ S^{\circ\prime})^{-1}E(S^\circ Y^\circ)}{E(D^{\circ 2}) - E(D^\circ S^{\circ\prime})E(S^\circ S^{\circ\prime})^{-1}E(S^\circ D^\circ)}. \qquad \text{(S1)}$$

Substituting the moments above and using $I - (\Sigma_\eta + \Omega)^{-1}\Sigma_\eta = (\Sigma_\eta + \Omega)^{-1}\Omega$ yields equation (9). Subtracting $\Omega$ from $E(S^\circ S^{\circ\prime})$ turns the latent block into $\Sigma_\eta$, and the first coefficient returns to $\theta_0$. Under the same residualization, the coefficient obtained without the latent covariates has bias $\gamma_0' \Sigma_\eta \beta_0 / C_{DD}$; dividing equation (9) by this quantity expresses the loss of adjustment in relative terms. Figure S1 shows both quantities along one path in which the proportion $R^2$ of latent variance explained by the observed covariates is varied jointly for the three factors; along other paths the relation need not be monotone.

The identity also holds at the sample moments. With $\hat{C} = N^{-1}\sum_i \widehat{W}_i^\circ \widehat{W}_i^{\circ\prime}$ and $\bar{\bar{\Omega}}_W = N^{-1}\sum_i \hat{\Omega}_{W,i}$,

$$\hat{a}_N = \hat{C}^{-1}\hat{b},\qquad \hat{a}_C = (\hat{C} - \bar{\bar{\Omega}}_W)^{-1}\hat{b},\qquad \hat{a}_N - \hat{a}_C = -\hat{C}^{-1}\bar{\bar{\Omega}}_W \hat{a}_C, \qquad (3)$$

so substituting $\hat{\Sigma}_\eta = \hat{C}_{SS} - \bar{\bar{\Omega}}$, $\hat{\gamma} = \hat{\Sigma}_\eta^{-1}\hat{C}_{SD}$, $\hat{\beta}_C$, and $\hat{C}_{DD}$ into equation (9) reproduces the difference between the uncorrected and corrected estimates exactly. This decomposes the correction into the residualized reliabilities, $\hat{\gamma}$, and $\hat{\beta}_C$ (Section 6 of the main text); it is not an independent prediction.

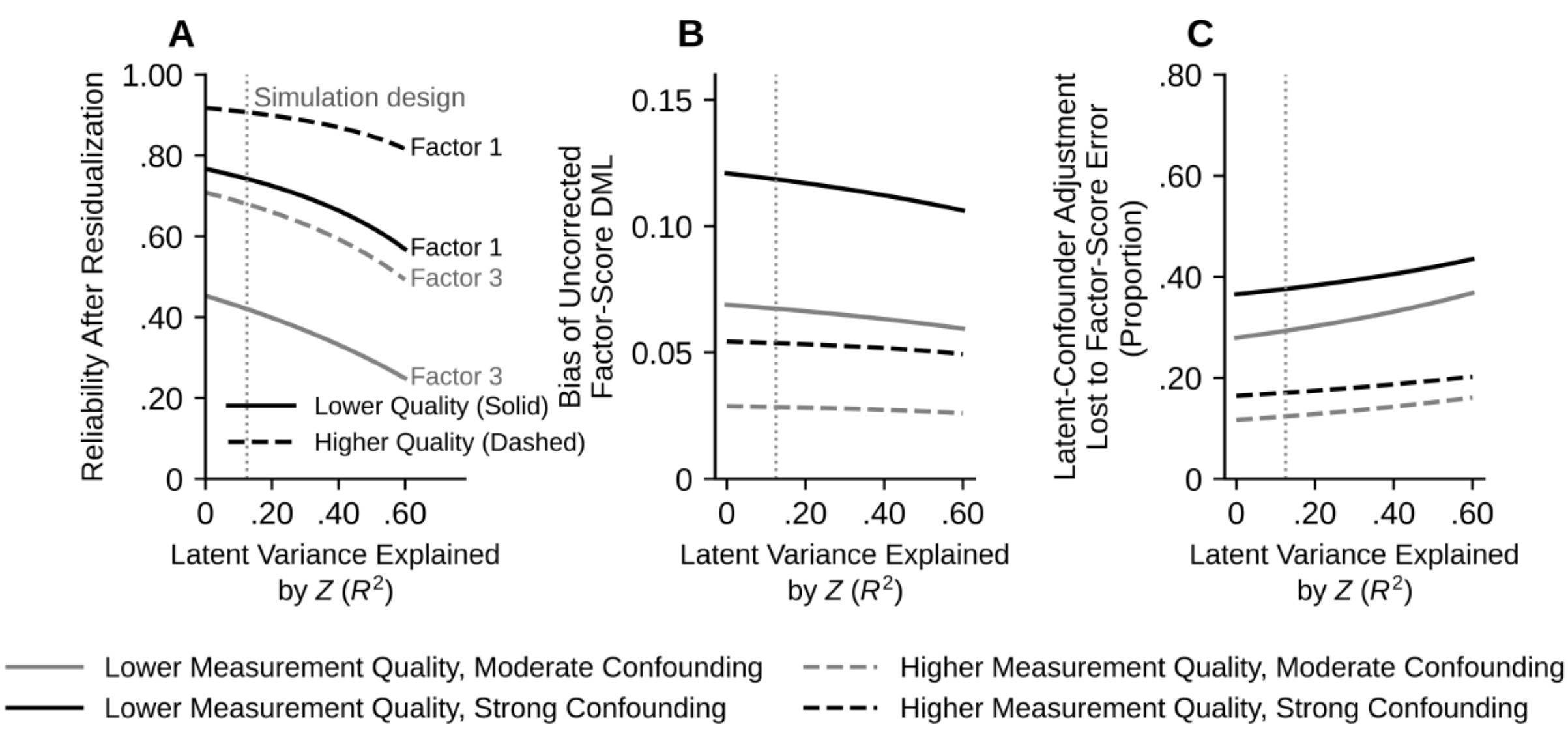

**Figure S1.** (A) Residualized reliability $\Sigma_{\eta,jj}/(\Sigma_{\eta,jj} + \Omega_{jj})$ of factors 1 and 3, (B) analytic bias of uncorrected factor-score DML from equation (9), and (C) proportion of the latent-covariate

adjustment lost to factor-score error (equation (9) divided by $\gamma_0'\Sigma_\eta\beta_0/C_{DD}$), as functions of the proportion $R^2$ of latent variance explained by the observed covariates. The two levels of measurement quality and the two confounding strengths of Section 4 of the main text are used, with latent correlations of .30 and $\mathrm{Var}(v) = 1$. The vertical dotted line marks the $R^2 = .125$ of the simulation design. DML = double/debiased machine learning.

***S1.2 Identification under the corrected estimating equation (Theorem 1)***

By equation (7) and A2, $E(S^\circ S^{\circ\prime}) = \Sigma_\eta + \Omega$ and $E(S^\circ D^\circ) = E(\eta^\circ D^\circ)$, so $A_0 = E(W^\circ W^{\circ\prime}) - \Omega_W$ equals the covariance matrix of $(D^\circ, \eta^{\circ\prime})'$. By equation (2), for any scalar $t$ and vector $r$,

$$(t, r')A_0(t, r')' = (r + t\gamma_0)'\Sigma_\eta(r + t\gamma_0) + t^2\mathrm{Var}(v), \qquad (4)$$

which is positive whenever $(t, r')' \neq 0$, because $\Sigma_\eta$ is positive definite and $\mathrm{Var}(v) > 0$ under A3. Hence $A_0$ is positive definite. At the true parameters, $(0, -\beta_0'\Omega)'$ in equation (8) and $\Omega_W a_0 = (0, \beta_0'\Omega)'$ cancel, so $E\{g(a_0)\} = 0$, and because $A_0$ is invertible, $a_0$ is the unique solution.

***S1.3 Neyman orthogonality with respect to the conditional means (Theorem 2)***

Let $\Delta\ell(Z), \Delta m(Z), \Delta q(Z)$ denote the directions of small perturbations of $\ell_0, m_0, q_0$, and let $e_1 = (1,0,\ldots,0)'$. At the true parameters, $R_0 = e - U'\beta_0$, $E(W^\circ \mid Z) = 0$, and $E(R_0 \mid Z) = 0$. The derivative in the $\ell$ direction is $-E\{W^\circ\Delta\ell(Z)\} = 0$. In the $m$ direction, $D^\circ$ changes by $-\Delta m(Z)$, so the derivative is $E\{-e_1\Delta m(Z)R_0 + W^\circ\theta_0\Delta m(Z)\} = 0$; in the $q$ direction, $S^\circ$ changes by $-\Delta q(Z)$, so the derivative is $E[-(0, \Delta q(Z)')'R_0 + W^\circ\Delta q(Z)'\beta_0] = 0$. Equation (10) is therefore Neyman orthogonal with respect to all three conditional means. If $\Omega$ is replaced by $\Omega + t\,\Delta\Omega$ with all other quantities held fixed, then $\partial_t E\{g(a_0; \Omega + t\,\Delta\Omega)\}|_{t=0} = \mathrm{blockdiag}(0, \Delta\Omega)a_0 = (0, (\Delta\Omega\,\beta_0)')'$, which is nonzero in general.

***S1.4 Consistency and asymptotic normality of the point estimator (Theorems 3–4)***

For a held-out observation $i \in I_k$, let $S_i = B(X_i - \tau)$ be the true Bartlett score and $\hat{S}_{ik} = \hat{B}_k(X_i - \hat{\tau}_k)$ the estimated score. Because $\hat{B}_k \to_p B$ and $\hat{\tau}_k \to_p \tau$ in each training fold and $E \parallel X \parallel^2 < \infty$, for fixed $K$,

$$N^{-1}\sum_i \parallel \hat{S}_i - S_i \parallel^2 = o_p(1). \qquad \text{(S2)}$$

Because $\hat{\ell}_k, \hat{m}_k, \hat{q}_k$ converge in mean square to $\ell_0, m_0, q_0$, in the cross-fitted held-out samples

$$N^{-1}\sum_i (\hat{Y}_i^\circ - Y_i^\circ)^2 = o_p(1), \qquad N^{-1}\sum_i \parallel \widehat{W}_i^\circ - W_i^\circ \parallel^2 = o_p(1). \qquad \text{(S3)}$$

Under cross-fitting, once the fitted results of each fold are held fixed, the held-out observations are independent of the training data, so a conditional law of large numbers applies. From equations (S2)–(S3) and the Cauchy–Schwarz inequality,

$$N^{-1}\sum_i \widehat{W}_i^\circ\,\widehat{W}_i^{\circ\prime} - N^{-1}\sum_i W_i^\circ\,W_i^{\circ\prime} = o_p(1), \qquad N^{-1}\sum_i \widehat{W}_i^\circ\,\hat{Y}_i^\circ - N^{-1}\sum_i W_i^\circ\,Y_i^\circ = o_p(1), \qquad \text{(S4)}$$

and, because $\widehat{\Omega}_k \to_p \Omega$ and $K$ is fixed,

$$N^{-1}\sum_i \widehat{\Omega}_{W,i} \to_p \Omega_W. \qquad \text{(S5)}$$

The law of large numbers applied to the cross-products of the true residuals gives

$$\hat{A} \to_p E(W^\circ W^{\circ\prime}) - \Omega_W = A_0, \qquad \hat{b} \to_p E(W^\circ Y^\circ) = A_0 a_0. \qquad \text{(S6)}$$

Because $\lambda_{\min}(A_0) > 0$, the probability that $\hat{A}$ is invertible tends to 1, and by the continuous mapping theorem $\hat{a} = \hat{A}^{-1}\hat{b} \to_p a_0$.

*Asymptotic normality (Theorem 4).* Write $\vartheta$ for the free measurement parameters (loadings, residual variances, factor correlations), $\hat{\vartheta}_k$ for their ML estimate in the training folds $I_k^c$, and $g_i(a; \eta, \vartheta)$ for the individual estimating function of equation (10) evaluated with conditional means $\eta = (\ell, m, q)$ and measurement parameters $\vartheta$, so that $S_i^\circ(\vartheta) = B(\vartheta)\{X_i - E(X_i \mid Z_i)\}$ and $\Omega(\vartheta) = B(\vartheta)\Theta(\vartheta)B(\vartheta)'$. Because $\hat{g}(a) = N^{-1}\sum_i g_i\,(a; \hat{\eta}_{k(i)}, \hat{\vartheta}_{k(i)}) = \hat{b} - \hat{A}a$ is linear in $a$, $\hat{a} - a_0 = \hat{A}^{-1}\hat{g}(a_0)$ exactly, and $\hat{A} \to_p A_0$ by Theorem 3. For each fold $k$,

$$\sum_{i\in I_k} g_i\,(a_0; \hat{\eta}_k, \hat{\vartheta}_k)$$
$$= \sum_{i\in I_k} g_i\,(a_0; \eta_0, \vartheta_0) + \sum_{i\in I_k}\{\, g_i(a_0; \hat{\eta}_k, \vartheta_0) - g_i(a_0; \eta_0, \vartheta_0)\}$$
$$+ \sum_{i\in I_k}\{\, g_i(a_0; \hat{\eta}_k, \hat{\vartheta}_k) - g_i(a_0; \hat{\eta}_k, \vartheta_0)\}, \qquad \text{(10)}$$

and we treat the three sums in turn.

(i) *Conditional means.* Conditionally on the training folds, the held-out observations are independent of $\hat{\eta}_k$. By Theorem 2, the terms of the second sum that are linear in $\hat{\ell}_k - \ell_0, \widehat{m}_k - m_0, \hat{q}_k - q_0$ have conditional mean zero, and their conditional variance is bounded by a constant times $|I_k|\ \|\ \hat{\eta}_k - \eta_0\ \|_2^2$, so they are $o_p(\sqrt{N})$ once $\|\ \hat{\eta}_k - \eta_0\ \|_2 = o_p(1)$. The conditional mean of the second sum consists of the second-order products $(\hat{\ell}_k - \ell_0)(\widehat{m}_k - m_0)$, $(\hat{\ell}_k - \ell_0)(\hat{q}_k - q_0)$, $(\widehat{m}_k - m_0)(\hat{q}_k - q_0)$, and $(\hat{q}_k - q_0)^2$ (the last is the $\Delta_q$ of Proposition 1), each of order $|I_k|$ times the product of the $L_2$ errors and hence $o_p(\sqrt{N})$ under the rate condition of Theorem 4 (the argument of Chernozhukov et al., 2018, Theorem 3.1). The condition holds when each learner converges at rate $o_p(N^{-1/4})$; for the correctly specified polynomial ridge learner it holds when the penalty is $o(N)$.

(ii) *Measurement parameters.* $\hat{\vartheta}_k$ depends only on the training folds, so the same conditioning argument and a Taylor expansion in $\vartheta$ give

$$\sum_{i\in I_k}\{\, g_i(a_0; \hat{\eta}_k, \hat{\vartheta}_k) - g_i(a_0; \hat{\eta}_k, \vartheta_0)\} = |I_k|\, J_\vartheta(\hat{\vartheta}_k - \vartheta_0) + o_p(\sqrt{N}), \qquad J_\vartheta$$
$$= \frac{\partial E\{g(a_0; \eta_0, \vartheta)\}}{\partial \vartheta'}\Big|_{\vartheta_0}, \qquad \text{(11)}$$

where $J_\vartheta$ collects the dependence through $B(\vartheta)$ in the residualized scores and through $\Omega(\vartheta)$ in the correction term. With the asymptotically linear representation $\hat{\vartheta}_k - \vartheta_0 = |I_k^c|^{-1}\sum_{j\in I_k^c}\psi\,(X_j) + o_p(N^{-1/2})$ and $|I_k|/|I_k^c| = 1/(K-1)$, summing over the $K$ folds gives $\sum_k |\, I_k| J_\vartheta(\hat{\vartheta}_k - \vartheta_0) = J_\vartheta \sum_j \psi\,(X_j) + o_p(\sqrt{N})$, because every observation belongs to $K-1$ training sets; the fold-wise estimators therefore have the same first-order representation as a full-sample estimator.

(iii) Collecting terms,

$$\sqrt{N}(\hat{a}-a_0) = A_0^{-1}N^{-1/2}\sum_i \{g_i(a_0;\eta_0,\vartheta_0) + J_\vartheta \psi(X_i)\} + o_p(1), \qquad (12)$$

a normalized sum of independent and identically distributed mean-zero terms with finite variance under the fourth-moment condition, so the central limit theorem and Slutsky's lemma give Theorem 4 with $V = \mathrm{Var}\{g_i(a_0) + J_\vartheta \psi(X_i)\}$. Because the subtracted covariance tracks the Bartlett weights, rescaling the loadings changes $\hat{\beta}_C$ but not $\hat{\theta}_C$, so the first-order measurement contribution to $\mathrm{Var}(\hat{\theta}_C)$ comes mainly from the residual variances.

*Stacked sandwich SE.* The sample analogue replaces $g_i(a_0)$ by the cross-fitted $\hat{g}_i(\hat{a})$, $J_\vartheta$ by the numerical derivative of the fold-$k$ moment sum with respect to $\hat{\vartheta}_k$, and $\psi(X_j)$ by the per-observation ML scores of the training-fold measurement model multiplied by the inverse observed information, $\hat{\mathcal{K}}_k$ (stacked estimating equations; Stefanski & Boos, 2002). With $u_i = \hat{g}_i(\hat{a}) + \sum_{k:\, i\in I_k^c} \hat{J}_k\, \hat{\mathcal{K}}_k \hat{\psi}_i^{(k)}$,

$$\widehat{\mathrm{Var}}(\hat{a}) = \hat{A}^{-1}\left\{N^{-2}\sum_i (u_i-\bar{u})(u_i-\bar{u})'\right\}\hat{A}^{-1}. \qquad (13)$$

The conditional sandwich SE omits the second term of $u_i$. Both treat the smoother weights and the selected penalties as fixed, which the orthogonality in (i) justifies to first order; for Croon-DML-R, $\hat{\Omega}_k \circ (1+G_i)$ replaces $\hat{\Omega}_k$ throughout. Section S9 compares the two SEs with the empirical standard deviation (Table S19).

### *S1.5 Bias of the population solution due to learning error in $q_0$ (Proposition 1)*

Suppose that the measurement model and $\ell_0, m_0$ are known and that only $q_0$ is estimated. In the held-out sample, let $\delta_q(Z) = \hat{q}(Z) - q_0(Z)$, so that $\hat{S}^\circ = S - \hat{q}(Z) = \eta^\circ + U - \delta_q(Z)$. Once the fitted results of a training–held-out split are held fixed, $\delta_q(Z)$ depends only on the $Z$ of the held-out observation, and $E(\eta^\circ + U \mid Z) = 0$; hence

$$E(\hat{S}^\circ \hat{S}^{\circ\prime} \mid \text{training sample}) = \Sigma_\eta + \Omega + \Delta_q, \qquad \Delta_q = E\{\delta_q(Z)\delta_q(Z)' \mid \text{training sample}\}. \qquad (14)$$

For the same reason, $E(\hat{S}^\circ D^\circ \mid \text{training sample}) = \Sigma_\eta \gamma_0$ and $E(\hat{S}^\circ Y^\circ \mid \text{training sample}) = \Sigma_\eta(\theta_0\gamma_0 + \beta_0)$, and the calculation of Section S1.1 gives

$$\theta_{C,q} - \theta_0 = \frac{\gamma_0'\Sigma_\eta(\Sigma_\eta + \Delta_q)^{-1}\Delta_q\beta_0}{C_{DD} - \gamma_0'\Sigma_\eta(\Sigma_\eta+\Delta_q)^{-1}\Sigma_\eta\gamma_0}, \qquad (\mathrm{S7})$$

where $\theta_{C,q}$ solves the estimating equation at the population moments of the held-out observations given the training sample. $\Delta_q$ is a conditional second moment and equals a covariance only when the conditional mean of the learning error is zero. If $\| \hat{q} - q_0 \|_2 = O_p(r_N)$, then $\| \Delta_q \| = O_p(r_N^2)$, and as long as the denominator is bounded away from zero, the bias is also $O_p(r_N^2)$.

### *S1.6 Noise fitting under a common prespecified smoother (Propositions 2–3; equations (14)–(15))*

This section corresponds to conditions C1–C3 of the main text. Suppose that the measurement parameters are known, that the rows are independent and identically distributed, that the smoother weights depend only on the training and evaluation predictors $Z$, and that, in addition to $E(U \mid \eta, Z) = 0$, $E(UU' \mid Z) = \Omega$ is constant. Writing the prediction for factor-score component $j$ as $\hat{q}_j(z) = w_j(z)'S_{j,\text{train}}$,

$$\delta_{q,j}(z) = \underbrace{w_j(z)'\eta_{j,\text{train}} - q_{0j}(z)}_{\delta_{\eta,j}(z)} + \underbrace{w_j(z)'U_{j,\text{train}}}_{\delta_{U,j}(z)}. \qquad (16)$$

With expectations taken over the training sample and an independent held-out observation, the second moment of the noise-fitting part is $E\{\delta_{U,j}(Z)\delta_{U,k}(Z)\} = \bar{G}_{jk}\Omega_{jk}$ with $\bar{G}_{jk} = E\{w_j(Z)'w_k(Z)\}$, so that

$$\Delta_U = E(\delta_U\delta_U') = \bar{G} \circ \Omega, \qquad \text{(S8)}$$

where $\circ$ denotes the Hadamard product. If the conditional error variance varies with $Z$, the error covariances of the individual training points enter the weighted sum and equation (S8) no longer separates. The cross-terms between signal error and noise error that $\Delta_q$ of Section S1.5 may contain vanish when the learning error is also averaged over, giving $E_{\text{train}}(\Delta_q) = E(\delta_\eta\delta_\eta') + \Delta_U$. If all factor-score components use the same prespecified smoother $w$, then $\Delta_U = \bar{\omega}\Omega$ with $\bar{\omega} = E\{\| w(Z) \|^2\}$, the mean sum of squared smoother weights at evaluation points (not the leverage of the training points); for ridge regression it is computed from the predictor matrix and the penalty parameter (Section S4.1).

If the three conditional means are estimated with the same prespecified $w$ (C1), the structural relation

$$\hat{Y}_i^\circ = \theta_0\hat{D}_i^\circ + \beta_0'\hat{S}_i^\circ - \beta_0'(U_i - w_i'U_{\text{train}}) + \{h_{Y0}(Z_i) - w_i'h_{Y0}(Z_{\text{train}})\} + (e_i - w_i'e_{\text{train}}) \qquad (18)$$

holds exactly: the signal learning error in $q_0$ enters $\hat{Y}_i^\circ$ through $\beta_0$ and cancels in the estimation of $\theta_0$, leaving products of the unremoved structural-function signals (second-order terms that the oracle also has). If the conditional second moments of $(\eta^\circ, v, e, U)$ do not vary with $Z$ (C2), subtracting a weighted sum of independent training-row noise from the noise of a held-out row multiplies its second moment by $1+\| w_i \|^2$, and averaging over held-out rows and training designs gives the common factor $1 + \bar{\omega}$. If the second moments and cross-moments of the unremoved signals are negligible (C3), the noise component of the corrected latent block becomes

$$(1 + \bar{\omega})(\Sigma_\eta + \Omega) - \Omega = (1 + \bar{\omega})\left\{\Sigma_\eta + \frac{\bar{\omega}}{1 + \bar{\omega}}\Omega\right\}. \qquad (19)$$

Dividing the other blocks of the normal equations by the same factor and applying equation (9) to the resulting moments yields

$$\theta_C - \theta_0 \approx \frac{\gamma_0'\Sigma_\eta(\Sigma_\eta + \Delta_U^*)^{-1}\Delta_U^*\beta_0}{C_{DD} - \gamma_0'\Sigma_\eta(\Sigma_\eta + \Delta_U^*)^{-1}\Sigma_\eta\gamma_0}, \qquad \Delta_U^* = \frac{\bar{\omega}}{1 + \bar{\omega}}\Omega, \qquad (20)$$

which is equation (14) and Proposition 2 of the main text. The approximation ignores the unremoved signals of $q_0, h_{D0}, h_{Y0}$, the estimation error in the measurement parameters, and finite-sample effects of the matrix inverse. For the uncorrected estimator no correction term is subtracted, so the common inflation of the noise part cancels.

For Croon-DML-R (Proposition 3 and equation (15) of the main text), the noise of held-out row $i$ is independent of the training rows, so its residualized factor-score noise $\widetilde{U}_i = U_i - w_i'U_{\text{train}}$ satisfies

$$E(\widetilde{U}_i\widetilde{U}_i' \mid Z) = \Omega \circ (1 + G_i), \qquad (G_i)_{jk} = w_j(Z_i)'w_k(Z_i). \qquad \text{(S9)}$$

Croon-DML-R subtracts exactly this quantity, so the noise component becomes $(1 + \bar{\omega})(\Sigma_\eta + \Omega) - (1 + \bar{\omega})\Omega = (1 + \bar{\omega})\Sigma_\eta$, inflated by the same factor as the $D^\circ$ block and the cross blocks; under C1–C3 the population solution is therefore $a_0$. Equation (S9) holds even when the penalties differ across components, so the noise component is removed exactly in that case too. Under target-specific LOOCV the weights depend on the target values, and the correction becomes an approximation that treats the selected penalties as fixed (Section S2.5). Signal learning error and the mismatch between target-specific smoothers are not targeted by this correction; they are addressed through the learner (Section S7).

## S2. Error sources, assumption violation, and misspecification

### *S2.1 Comparison of error sources*

Five settings are compared at $N = 1000$ and 5000 with 1,000 replications per condition (Table S1): an oracle with the true conditional means; an uncorrected estimator with Bartlett scores from the true measurement model and the true conditional means; the same setting with the Ω correction; Croon-DML with the true measurement model but learned conditional means; and Croon-DML with both estimated. The sample means of the uncorrected estimating equation at the true parameters agreed with the values $(0, -0.122, -0.182, -0.242)'$ of equation (8) within Monte Carlo error, and the means of the corrected estimating equation were close to zero. In a separate setting, only $q_0$ is learned by cross-fitting at $N = 500, 1000, 2000, 5000$ with 1,000 replications; in each replication, $\widehat{\Delta}_q = N^{-1} \sum_i \delta_q(Z_i)\delta_q(Z_i)'$ is computed and the observed bias is compared with the prediction of equation (S7) (Table S2).

**Table S1.** Comparison of error sources (lower measurement quality, strong confounding, $\theta_0 = 0.30$, 1,000 replications per condition)

| $N$ | Setting | Bias | MCSE | Empirical SD | RMSE |
|---|---|---|---|---|---|
| 1000 | Croon-DML | 0.022 | 0.001 | 0.038 | 0.044 |
| 1000 | Croon, true conditional means | −0.002 | 0.001 | 0.040 | 0.040 |
| 1000 | Uncorrected, true conditional means | 0.117 | 0.001 | 0.031 | 0.121 |
| 1000 | Oracle, true conditional means | −0.002 | 0.001 | 0.032 | 0.032 |
| 1000 | Croon-DML, true measurement model | 0.019 | 0.001 | 0.039 | 0.043 |
| 5000 | Croon-DML | 0.008 | 0.001 | 0.017 | 0.019 |
| 5000 | Croon, true conditional means | −0.001 | 0.001 | 0.017 | 0.017 |
| 5000 | Uncorrected, true conditional means | 0.118 | 0.000 | 0.013 | 0.119 |
| 5000 | Oracle, true conditional means | −0.000 | 0.000 | 0.014 | 0.014 |
| 5000 | Croon-DML, true measurement model | 0.007 | 0.001 | 0.017 | 0.019 |

*Note.* Bias = mean estimate − $\theta_0$; MCSE = Monte Carlo standard error of bias; SD = empirical standard deviation of the estimates; RMSE = root mean square error. The analytic bias from equation (9) is 0.119 at both sample sizes.

**Table S2.** Analytic prediction and empirical bias for learning error in $q_0$ (lower measurement quality, strong confounding, 1,000 replications per condition)

| $N$ | Croon bias, true conditional means | Croon-DML bias, only $q$ estimated (MCSE) | Prediction from equation (S7) | Difference | RMSE of $q$ prediction | $\mathrm{tr}(\widehat{\Delta}_q)$ |
|---|---|---|---|---|---|---|
| 500 | −0.005 | 0.038 (0.002) | 0.041 | −0.004 | 0.348 | 0.364 |
| 1000 | −0.002 | 0.035 (0.001) | 0.037 | −0.002 | 0.323 | 0.314 |
| 2000 | −0.001 | 0.031 (0.001) | 0.031 | −0.000 | 0.291 | 0.254 |
| 5000 | −0.001 | 0.021 (0.001) | 0.022 | −0.001 | 0.234 | 0.164 |

*Note.* MCSE = Monte Carlo standard error; Difference = observed bias − prediction from equation (S7); RMSE of $q$ prediction = $\sqrt{\mathrm{tr}(\widehat{\Delta}_q)/3}$, the root mean square error per component of the prediction of $q_0$ (the standard deviation of each component of $q_0$ is 0.354). Unlike the main design, only $q_0$ is learned here; when all three conditional means are learned, the shared learning errors partly cancel (Section S1.6).

### *S2.2 Violation of A2*

If $E(U \mid \eta, Z) = 0$ is maintained but $c_v = E(Uv)$ and $c_e = E(Ue)$ may be nonzero, the expectation of the corrected estimating equation at the true parameters is

$$E\{g(a_0)\} = (-\beta_0' c_v,\ c_e')'. \qquad \text{(S10)}$$

With $c = \Sigma_\eta \gamma_0 + c_v$, and provided that the relevant matrix is invertible, the bias of the population solution (pseudo-true value) of the corrected estimating equation is

$$\theta_{C,A2} - \theta_0 = \frac{-c_v'\beta_0 - c'\Sigma_\eta^{-1} c_e}{C_{DD} - c'\Sigma_\eta^{-1} c}. \qquad \text{(S11)}$$

In the simulation, the structural error of the focal-predictor equation is $v = \sqrt{1 - \kappa^2 \sigma_h^2}\, v_0 + \kappa h$ with $h = \varepsilon_1 + \varepsilon_5 + \varepsilon_9$, which keeps $\mathrm{Var}(v) = 1$; then $c_e = 0$ and $c_v = \kappa B \Theta s$, where $s$ selects $\varepsilon_1, \varepsilon_5, \varepsilon_9$. For $\kappa = 0, 0.10, 0.20, 0.30, 0.40$ and $N = 1000, 5000$ (1,000 replications per condition), $U = S - \eta$ and $v$ are recovered from the generating quantities to compute $\hat{c}_v = N^{-1} \sum_i U_i\, v_i$ (Table S3). The uncorrected bias decreased with $\kappa$ because the negative distortion due to the violation partly offset the positive bias due to factor-score error.

**Table S3.** Covariances and bias in the structural coefficient by strength of the A2 violation (lower measurement quality, strong confounding, 1,000 replications per condition)

| $N$ | $\kappa$ | Population $\lVert c_v \rVert$ | Sample $\lVert \hat{c}_v \rVert$ | Prediction from equation (S11) | Croon-DML bias (MCSE) | Uncorrected bias |
|---|---|---|---|---|---|---|
| 1000 | 0.00 | 0.000 | 0.042 | −0.000 | 0.022 (0.001) | 0.113 |
| 1000 | 0.10 | 0.064 | 0.073 | −0.031 | −0.000 (0.001) | 0.104 |
| 1000 | 0.20 | 0.129 | 0.134 | −0.069 | −0.025 (0.001) | 0.095 |
| 1000 | 0.30 | 0.193 | 0.197 | −0.117 | −0.056 (0.002) | 0.083 |
| 1000 | 0.40 | 0.258 | 0.261 | −0.181 | −0.095 (0.002) | 0.071 |
| 5000 | 0.00 | 0.000 | 0.019 | −0.000 | 0.008 (0.001) | 0.118 |
| 5000 | 0.10 | 0.064 | 0.066 | −0.031 | −0.020 (0.001) | 0.108 |

| $N$ | $\kappa$ | Population ‖ $c_v$ ‖ | Sample ‖ $\hat{c}_v$ ‖ | Prediction from equation (S11) | Croon-DML bias (MCSE) | Uncorrected bias |
|---|---|---|---|---|---|---|
| 5000 | 0.20 | 0.129 | 0.130 | −0.069 | −0.055 (0.001) | 0.095 |
| 5000 | 0.30 | 0.193 | 0.194 | −0.117 | −0.099 (0.001) | 0.082 |
| 5000 | 0.40 | 0.258 | 0.258 | −0.181 | −0.153 (0.001) | 0.068 |

*Note.* $\kappa$ = violation strength; $c_v = E(Uv)$ = covariance between the factor-score error and the structural error of the focal-predictor equation; MCSE = Monte Carlo standard error. Sample ‖ $\hat{c}_v$ ‖ is the mean of the per-replication norms and is positive at $\kappa = 0$ because of sampling error. Equation (S11) is the solution at the true population moments, whereas the observed Croon-DML bias also includes the residual bias of a procedure in which the measurement parameters and conditional means are estimated (Section 5.2 of the main text).

### *S2.3 Misspecification of the conditional means*

At $N = 1000{,}5000$, five settings are compared on the same generated data with 500 replications per condition (Table S4). In the correct specification, $\ell_0, m_0, q_0$ are fitted by quadratic polynomial ridge regression. Fitting $\ell_0$ or $m_0$ by linear ridge regression removes the quadratic and interaction terms from that conditional mean, which corresponds to adjusting for the observed covariates linearly. Because $q_0$ is linear in the data generation, its misspecification is represented by omitting its predictors $Z_1$–$Z_6$; joint misspecification applies the three errors together.

**Table S4.** Misspecification of individual conditional means (lower measurement quality, strong confounding, 500 replications per condition)

| $N$ | Conditional-mean specification | Oracle bias | Uncorrected bias | Croon bias | Uncorrected − oracle | Croon − oracle |
|---|---|---|---|---|---|---|
| 1000 | Correct specification | 0.007 | 0.112 | 0.021 | 0.105 | 0.014 |
| 1000 | Linear ridge for $\ell_0$ only | −0.083 | 0.020 | −0.074 | 0.103 | 0.008 |
| 1000 | Linear ridge for $m_0$ only | −0.134 | −0.042 | −0.125 | 0.092 | 0.008 |
| 1000 | $q_0$ predictors omitted | 0.035 | 0.121 | 0.038 | 0.086 | 0.003 |
| 1000 | Joint misspecification | −0.102 | −0.026 | −0.100 | 0.076 | 0.002 |
| 5000 | Correct specification | 0.002 | 0.118 | 0.008 | 0.117 | 0.007 |
| 5000 | Linear ridge for $\ell_0$ only | −0.021 | 0.092 | −0.017 | 0.113 | 0.004 |
| 5000 | Linear ridge for $m_0$ only | −0.124 | −0.035 | −0.120 | 0.089 | 0.004 |
| 5000 | $q_0$ predictors omitted | 0.040 | 0.134 | 0.040 | 0.094 | 0.001 |
| 5000 | Joint misspecification | −0.098 | −0.023 | −0.098 | 0.075 | 0.000 |

*Note.* Oracle = oracle latent-variable DML; uncorrected = uncorrected factor-score DML; Croon = Croon-DML; DML = double/debiased machine learning. The difference columns give each estimator's bias minus the oracle's bias on the same generated data.

***S2.4 Misspecification of the measurement model***

Error correlations of .30 are generated for the indicator pairs $x_4$–$x_5$ and $x_8$–$x_9$, and a measurement model that estimates the two error covariances is compared, on the same generated data, with one that fixes them at zero ($N = 1000{,}5000$; 500 replications per condition; Table S5). Because omitting the error covariances changes both $B$ and $\Omega$, this evaluates the sensitivity of a correction obtained from the measurement model (Cole et al., 2007). The estimation failure rate was .000.

**Table S5.** Misspecification of the measurement model (lower measurement quality, strong confounding, 500 replications per condition)

| $N$ | Measurement model | Oracle bias | Uncorrected bias | Croon bias | Croon RMSE | Increase in Croon bias |
|---|---|---|---|---|---|---|
| 1000 | Correct specification | 0.010 | 0.119 | 0.023 | 0.046 | — |
| 1000 | Error covariances omitted | 0.010 | 0.121 | 0.037 | 0.054 | 0.014 |
| 5000 | Correct specification | 0.001 | 0.124 | 0.009 | 0.020 | — |
| 5000 | Error covariances omitted | 0.001 | 0.126 | 0.027 | 0.032 | 0.018 |

*Note.* Oracle = oracle latent-variable DML; uncorrected = uncorrected factor-score DML; Croon = Croon-DML; RMSE = root mean square error; DML = double/debiased machine learning. Increase in Croon bias = difference in the mean Croon-DML estimate between the omitted and correct specifications on the same generated data.

***S2.5 Penalty parameter and noise-fit diagnostics***

With the seeds and folds of the first 250 replications of the main design, the penalty parameter of the quadratic polynomial ridge regression is changed from target-specific LOOCV (16 candidate values from $10^{-1}$ to $10^{4}$) to $\alpha = 1$ for all targets. Because $Y$, $D$, and the three factor-score components then share one prespecified linear smoother, this condition evaluates equation (S8) and the approximation of equation (14). Because $U = S - \eta$ is known in the simulation, the folds and penalties actually used by the $q_0$ learner are applied to the training-fold $U$, and $\widehat{\Delta}_U$ is computed from the held-out fitted values (the empirical noise-fit ratio $\mathrm{tr}(\widehat{\Delta}_U)/\mathrm{tr}(\Omega)$). For ridge regression, the design-based ratio from $\bar{G} \circ \Omega$ (Section S4.1) is also computed with the same folds and penalties; the two ratios differed by at most 0.015 under $\alpha = 1$ and by 0.000 (after rounding) under LOOCV, which supports treating the selected penalties as fixed. Table S6 reports the paired performance under strong confounding (panel A) and the noise-fit diagnostics for all 12 conditions (panel B).

**Table S6.** Performance and noise-fit diagnostics by penalty parameter

**A. Paired performance under strong confounding**

| $N$ | Measurement quality | $\alpha = 1$ bias | $\alpha = 1$ RMSE | LOOCV bias | LOOCV RMSE | Noise-fit ratio: $\alpha = 1$ | Noise-fit ratio: LOOCV |
|---|---|---|---|---|---|---|---|
| 500 | Lower | 0.090 | 0.120 | 0.023 | 0.059 | 1.63 | 0.014 |

| $N$ | Measurement quality | $\alpha = 1$ bias | $\alpha = 1$ RMSE | LOOCV bias | LOOCV RMSE | Noise-fit ratio: $\alpha = 1$ | Noise-fit ratio: LOOCV |
|---|---|---|---|---|---|---|---|
| 500 | Higher | 0.036 | 0.090 | 0.016 | 0.057 | 1.65 | 0.019 |
| 1000 | Lower | 0.049 | 0.066 | 0.021 | 0.044 | 0.45 | 0.014 |
| 1000 | Higher | 0.021 | 0.051 | 0.015 | 0.039 | 0.45 | 0.021 |
| 5000 | Lower | 0.012 | 0.020 | 0.008 | 0.018 | 0.06 | 0.018 |
| 5000 | Higher | 0.005 | 0.017 | 0.005 | 0.017 | 0.06 | 0.025 |

**B. Noise-fit diagnostics for the 12 main conditions**

| $N$ | Measurement quality | Confounding | Mean sum of squared weights $\bar{\omega}$ ($\alpha = 1$) | Noise-fit ratio ($\alpha = 1$) | Bias approximation from equation (14) ($\alpha = 1$) | Croon − oracle ($\alpha = 1$) | Selected $\alpha$ (LOOCV) | Noise-fit ratio (LOOCV) | Croon − oracle (LOOCV) |
|---|---|---|---|---|---|---|---|---|---|
| 500 | Lower | Moderate | 1.64 | 1.63 | 0.048 | 0.047 | 3247 | 0.013 | 0.005 |
| 500 | Lower | Strong | 1.64 | 1.63 | 0.087 | 0.094 | 3217 | 0.014 | 0.011 |
| 500 | Higher | Moderate | 1.64 | 1.62 | 0.019 | 0.019 | 2394 | 0.019 | 0.002 |
| 500 | Higher | Strong | 1.64 | 1.65 | 0.036 | 0.038 | 2426 | 0.019 | 0.004 |
| 1000 | Lower | Moderate | 0.45 | 0.45 | 0.027 | 0.027 | 3163 | 0.014 | 0.006 |
| 1000 | Lower | Strong | 0.45 | 0.45 | 0.051 | 0.052 | 3178 | 0.014 | 0.014 |
| 1000 | Higher | Moderate | 0.45 | 0.45 | 0.010 | 0.010 | 2207 | 0.021 | 0.003 |
| 1000 | Higher | Strong | 0.45 | 0.45 | 0.019 | 0.021 | 2223 | 0.021 | 0.005 |
| 5000 | Lower | Moderate | 0.06 | 0.06 | 0.006 | 0.006 | 2820 | 0.018 | 0.003 |
| 5000 | Lower | Strong | 0.06 | 0.06 | 0.012 | 0.012 | 2829 | 0.018 | 0.007 |
| 5000 | Higher | Moderate | 0.06 | 0.06 | 0.002 | 0.002 | 2126 | 0.024 | 0.001 |
| 5000 | Higher | Strong | 0.06 | 0.06 | 0.004 | 0.005 | 2130 | 0.025 | 0.003 |

*Note.* Noise-fit ratio = $\mathrm{tr}(\widehat{\Delta}_U)/\mathrm{tr}(\Omega)$, the variance of the part of the factor-score error $U$ that the $q_0$ learner fits as if it were signal, relative to the factor-score error variance; $\bar{\omega}$ = mean sum of squared smoother weights; $\alpha = 1$ = penalty parameter fixed at 1 for all targets (common prespecified smoother); LOOCV = penalty parameter selected separately for each target by leave-one-out cross-validation; RMSE = root mean square error. Panel A uses the first 250 paired replications; in panel B, the $\alpha = 1$ columns summarize 250 replications and the LOOCV columns 1,000. The bias approximation from equation (14) substitutes $\bar{\omega}$ into equation (14) of the main text and applies only to $\alpha = 1$. Croon − oracle is the bias difference between the two estimators in the same replications. Selected $\alpha$ is the geometric mean, over replications and folds, of the penalty for the standardized second-degree polynomial terms.

## S3. Nonlinear high-dimensional conditions and approximate intervals

### *S3.1 Nonlinear high-dimensional conditional means*

With 100 observed covariates and

$$f_D(Z) = 0.55\sin Z_1 + 0.35 Z_2 Z_3 + 0.30\tanh Z_4 + 0.25 I(Z_5 > 0) + 0.20|Z_6|, \quad (24)$$

$$f_Y(Z) = 0.50\sin(Z_2 Z_3) - 0.30|Z_4| + 0.30 Z_7 Z_8 + 0.25 I(Z_9 > 0.25) + 0.20\log(1 + Z_{10}^2), \quad (25)$$

data are generated at $N = 500, 1000, 5000$ with lower or higher measurement quality and strong confounding (500 replications per condition); $q_0$ is linear as in the main design. The learners are histogram-based gradient boosting (HGB), random forests, quadratic polynomial ridge regression with $\alpha = 1$ (5,150 second-degree terms), and quadratic polynomial ridge regression with LOOCV (settings in Section S4.1). The noise-fit ratio is recorded for LOOCV ridge regression, computed post hoc for $\alpha = 1$ on the first 100 replications per condition, and not computed for the tree learners, which are not linear smoothers. Under $\alpha = 1$, the ratio rises to 0.10, 0.20, and 3.72–3.74 at $N = 500, 1000, 5000$, because the prediction variance under a weak penalty grows as the number of terms approaches the training sample size (Hastie et al., 2022).

The geometric mean of the $\alpha$ selected by LOOCV for the factor-score targets was 8,883–10,000, at the upper limit of the candidate values. With the candidate range extended to $10^{-1}$–$10^{7}$ (25 values) on the same samples and folds (250 replications; the Ridge (CV+) rows of Table S7), the selected $\alpha$ was 49,863–217,437, the noise-fit ratio fell to 0.004–0.008, and the Croon − oracle difference at $N = 5000$ decreased from 0.004/0.011 (higher/lower quality) to 0.001/0.002; the bias and RMSE of Croon-DML changed by at most 0.004 and 0.003.

**Table S7.** Bias and RMSE with nonlinear high-dimensional conditional means (strong confounding; 500 replications per condition, 250 for the extended candidate range)

| $N$ | Measurement quality | Learner | Oracle bias | Uncorrected bias | Croon bias | Uncorrected RMSE | Croon RMSE | Noise-fit ratio | Selected $\alpha$ | Croon − oracle |
|---|---|---|---|---|---|---|---|---|---|---|
| 500 | Higher | HGB | 0.040 | 0.081 | 0.048 | 0.096 | 0.073 | — | — | 0.008 |
| 500 | Higher | Random forest | 0.052 | 0.096 | 0.055 | 0.107 | 0.074 | — | — | 0.003 |
| 500 | Higher | Ridge (fixed) | 0.045 | 0.089 | 0.051 | 0.102 | 0.074 | 0.10 | — | 0.006 |
| 500 | Higher | Ridge (CV) | 0.048 | 0.091 | 0.051 | 0.101 | 0.069 | 0.013 | 8,883 | 0.003 |
| 500 | Higher | Ridge (CV+) | 0.047 | 0.088 | 0.048 | 0.097 | 0.065 | 0.008 | 176,216 | 0.001 |
| 500 | Lower | HGB | 0.035 | 0.127 | 0.055 | 0.136 | 0.078 | — | — | 0.020 |
| 500 | Lower | Random forest | 0.049 | 0.147 | 0.055 | 0.153 | 0.076 | — | — | 0.006 |
| 500 | Lower | Ridge (fixed) | 0.042 | 0.142 | 0.060 | 0.149 | 0.082 | 0.10 | — | 0.018 |
| 500 | Lower | Ridge (CV) | 0.045 | 0.142 | 0.052 | 0.148 | 0.072 | 0.013 | 9,110 | 0.007 |
| 500 | Lower | Ridge (CV+) | 0.048 | 0.145 | 0.056 | 0.150 | 0.074 | 0.007 | 217,437 | 0.007 |
| 1000 | Higher | HGB | 0.030 | 0.071 | 0.035 | 0.079 | 0.050 | — | — | 0.005 |

| ***N*** | **Measurement quality** | **Learner** | **Oracle bias** | **Uncorrected bias** | **Croon bias** | **Uncorrected RMSE** | **Croon RMSE** | **Noise-fit ratio** | **Selected $\alpha$** | **Croon − oracle** |
|---|---|---|---|---|---|---|---|---|---|---|
| 1000 | Higher | Random forest | 0.044 | 0.088 | 0.045 | 0.093 | 0.056 | — | — | 0.001 |
| 1000 | Higher | Ridge (fixed) | 0.032 | 0.077 | 0.041 | 0.086 | 0.058 | 0.20 | — | 0.009 |
| 1000 | Higher | Ridge (CV) | 0.040 | 0.083 | 0.042 | 0.089 | 0.053 | 0.018 | 9,955 | 0.001 |
| 1000 | Higher | Ridge (CV+) | 0.039 | 0.081 | 0.039 | 0.086 | 0.050 | 0.005 | 133,263 | 0.000 |
| 1000 | Lower | HGB | 0.033 | 0.125 | 0.047 | 0.129 | 0.060 | — | — | 0.014 |
| 1000 | Lower | Random forest | 0.048 | 0.145 | 0.051 | 0.148 | 0.064 | — | — | 0.002 |
| 1000 | Lower | Ridge (fixed) | 0.036 | 0.137 | 0.063 | 0.142 | 0.078 | 0.20 | — | 0.027 |
| 1000 | Lower | Ridge (CV) | 0.044 | 0.141 | 0.048 | 0.144 | 0.062 | 0.018 | 9,967 | 0.004 |
| 1000 | Lower | Ridge (CV+) | 0.046 | 0.141 | 0.048 | 0.145 | 0.061 | 0.004 | 212,857 | 0.003 |
| 5000 | Higher | HGB | 0.011 | 0.057 | 0.015 | 0.060 | 0.022 | — | — | 0.004 |
| 5000 | Higher | Random forest | 0.034 | 0.079 | 0.035 | 0.081 | 0.039 | — | — | 0.001 |
| 5000 | Higher | Ridge (fixed) | 0.003 | 0.055 | 0.046 | 0.064 | 0.057 | 3.72 | — | 0.043 |
| 5000 | Higher | Ridge (CV) | 0.031 | 0.077 | 0.035 | 0.078 | 0.039 | 0.067 | 10,000 | 0.004 |
| 5000 | Higher | Ridge (CV+) | 0.037 | 0.081 | 0.038 | 0.082 | 0.041 | 0.007 | 49,863 | 0.001 |
| 5000 | Lower | HGB | 0.012 | 0.115 | 0.022 | 0.116 | 0.029 | — | — | 0.010 |
| 5000 | Lower | Random forest | 0.035 | 0.136 | 0.037 | 0.136 | 0.041 | — | — | 0.002 |
| 5000 | Lower | Ridge (fixed) | 0.004 | 0.120 | 0.104 | 0.124 | 0.109 | 3.74 | — | 0.100 |
| 5000 | Lower | Ridge (CV) | 0.032 | 0.133 | 0.043 | 0.134 | 0.046 | 0.067 | 10,000 | 0.011 |
| 5000 | Lower | Ridge (CV+) | 0.039 | 0.138 | 0.041 | 0.138 | 0.045 | 0.004 | 71,882 | 0.002 |

*Note.* HGB = histogram-based gradient boosting; Ridge (fixed) = quadratic polynomial ridge regression with $\alpha = 1$; Ridge (CV) = quadratic polynomial ridge regression with target-specific leave-one-out cross-validation over $10^{-1}$–$10^{4}$; Ridge (CV+) = the same over the extended range $10^{-1}$–$10^{7}$; oracle = oracle latent-variable DML; uncorrected = uncorrected factor-score DML; Croon = Croon-DML; RMSE = root mean square error; DML = double/debiased machine

learning. Noise-fit ratio = $\mathrm{tr}(\widehat{\Delta}_U)/\mathrm{tr}(\Omega)$ (not computed for the tree learners, —); Selected $\alpha$ = geometric mean over replications and folds for the factor-score targets. For Ridge (fixed), the bias approximations from equation (14) were 0.006/0.017 (higher/lower quality) at $N = 500$, 0.011/0.031 at $N = 1000$, and 0.044/0.102 at $N = 5000$, close to the Croon − oracle values of 0.006/0.018, 0.009/0.027, and 0.043/0.100. The estimation failure rate was .000.

### *S3.2 Approximate intervals from the fixed aggregated-penalty bootstrap*

In 12 conditions ($N = 500{,}1000{,}5000$ × lower/higher measurement quality × $\theta_0 = 0{,}0.30$; strong confounding), 500 outer Monte Carlo replications are run with 300 bootstrap resamples per sample. Point estimates in the original sample use penalties selected by LOOCV separately for each fold and target. In the resamples, the fold-specific penalties of the original sample are aggregated by their geometric mean for each target, and this value is held fixed in all resample folds (the fixed aggregated-penalty bootstrap); the empirical illustration uses the same rule (Section S8.1). In every resample, folds are formed anew from the unique original-sample indices so that all copies of an observation fall in one fold, and the measurement model, $\widehat{B}, \widehat{\Omega}$, the conditional means, and the structural coefficients are re-estimated. An interval is computed only if at least 90% of the re-estimations succeed.

Three 95% intervals are compared. The percentile interval uses the 2.5th and 97.5th percentiles of the resample estimates. The bias-corrected (BC) interval obtains $z_0 = \Phi^{-1}(p_0)$ from the proportion $p_0$ of resample estimates below the original point estimate (ties receive half weight; with $R$ valid resamples, $p_0$ is restricted to $[1/(2R), 1 - 1/(2R)]$) and uses the $\Phi(2z_0 \mp 1.96)$ quantiles of the resample distribution (Efron, 1987). The Wald interval is the point estimate $\pm 1.96$ conditional sandwich standard errors, computed with the measurement parameters and conditional means held fixed from $\hat{A}$ in equation (12) and the sample covariance of the corrected estimating equation. With 300 resamples the relative Monte Carlo error of the bootstrap SE is about 4%, and with 500 outer replications the MCSE of coverage is about 0.010.

In 14 of the 6,000 outer replications (samples with $N = 500$ and lower measurement quality, 7 per $\theta_0$), the re-estimation success rate fell below .90 for uncorrected DML and Croon-DML, so no intervals were computed for them. The ratio of bootstrap SE to empirical SD was 1.06–1.29 for all estimators, including the oracle, which involves no measurement model: placing the duplicated observations of a resample in the same fold reduces the effective training sample size and increases the learning error of the conditional means. The results are interpreted in Section 5.5 of the main text.

**Table S8.** Performance of approximate intervals from the fixed aggregated-penalty bootstrap at $\theta_0 = 0$ (strong confounding, 500 outer replications × 300 bootstrap resamples)

| *N* | Measurement quality | Method | Bias (MCSE) | Boot SE/SD | Sandwich SE/SD | Wald coverage | Percentile coverage | BC coverage | Width (Wald) | Width (BC) |
|---|---|---|---|---|---|---|---|---|---|---|
| 500 | Lower | Oracle | 0.015 (0.002) | 1.29 | — | — | .988 | .958 | — | 0.214 |
| 500 | Lower | Uncorrected | 0.111 (0.002) | 1.27 | 0.99 | .254 | .381 | .473 | 0.168 | 0.209 |
| 500 | Lower | Croon-DML | 0.027 (0.002) | 1.25 | 1.00 | .919 | .968 | .945 | 0.201 | 0.242 |

| $N$ | Measurement quality | Method | Bias (MCSE) | Boot SE/SD | Sandwich SE/SD | Wald coverage | Percentile coverage | BC coverage | Width (Wald) | Width (BC) |
|---|---|---|---|---|---|---|---|---|---|---|
| 500 | Higher | Oracle | 0.019 (0.002) | 1.24 | — | — | .982 | .942 | — | 0.216 |
| 500 | Higher | Uncorrected | 0.060 (0.002) | 1.22 | 0.95 | .708 | .868 | .818 | 0.171 | 0.213 |
| 500 | Higher | Croon-DML | 0.023 (0.002) | 1.23 | 0.97 | .918 | .982 | .946 | 0.182 | 0.224 |
| 1000 | Lower | Oracle | 0.012 (0.002) | 1.20 | — | — | .974 | .936 | — | 0.156 |
| 1000 | Lower | Uncorrected | 0.117 (0.001) | 1.22 | 0.92 | .040 | .062 | .166 | 0.118 | 0.152 |
| 1000 | Lower | Croon-DML | 0.024 (0.002) | 1.17 | 0.92 | .872 | .924 | .934 | 0.144 | 0.175 |
| 1000 | Higher | Oracle | 0.015 (0.002) | 1.21 | — | — | .968 | .952 | — | 0.156 |
| 1000 | Higher | Uncorrected | 0.061 (0.002) | 1.22 | 0.92 | .500 | .612 | .688 | 0.121 | 0.155 |
| 1000 | Higher | Croon-DML | 0.020 (0.002) | 1.20 | 0.92 | .878 | .944 | .936 | 0.130 | 0.162 |
| 5000 | Lower | Oracle | 0.002 (0.001) | 1.06 | — | — | .962 | .942 | — | 0.064 |
| 5000 | Lower | Uncorrected | 0.119 (0.001) | 1.09 | 0.93 | .000 | .000 | .000 | 0.053 | 0.061 |
| 5000 | Lower | Croon-DML | 0.009 (0.001) | 1.07 | 0.93 | .894 | .894 | .936 | 0.067 | 0.074 |
| 5000 | Higher | Oracle | 0.002 (0.001) | 1.16 | — | — | .980 | .962 | — | 0.064 |
| 5000 | Higher | Uncorrected | 0.055 (0.001) | 1.17 | 0.99 | .026 | .036 | .072 | 0.054 | 0.063 |
| 5000 | Higher | Croon-DML | 0.005 (0.001) | 1.15 | 0.98 | .936 | .960 | .952 | 0.059 | 0.067 |

*Note.* Oracle = oracle latent-variable DML; uncorrected = uncorrected factor-score DML; MCSE = Monte Carlo standard error; Boot SE/SD = mean bootstrap standard error divided by the empirical standard deviation; Sandwich SE/SD = mean conditional sandwich standard error divided by the empirical standard deviation (not applicable to the oracle); BC = bias-corrected interval; Width = mean interval width; DML = double/debiased machine learning. Coverage is the proportion of intervals containing $\theta_0$; because $\theta_0 = 0$, the Type I error rate is $1 - \text{coverage}$. For uncorrected DML and Croon-DML at $N = 500$ with lower measurement quality, results are based on the 493 of 500 replications without interval estimation failures.

**Table S9.** Performance of approximate intervals from the fixed aggregated-penalty bootstrap at $\theta_0 = 0.30$ (strong confounding, 500 outer replications × 300 bootstrap resamples)

| $N$ | Measurement quality | Method | Bias (MCSE) | Boot SE/SD | Sandwich SE/SD | Wald coverage | Percentile coverage | BC coverage | Width (Wald) | Width (BC) |
|---|---|---|---|---|---|---|---|---|---|---|
| 500 | Lower | Oracle | 0.011 (0.002) | 1.21 | — | — | .982 | .934 | — | 0.211 |
| 500 | Lower | Uncorrected | 0.107 (0.002) | 1.22 | 0.96 | .286 | .383 | .505 | 0.168 | 0.206 |
| 500 | Lower | Croon-DML | 0.024 (0.002) | 1.19 | 0.96 | .907 | .959 | .923 | 0.200 | 0.239 |
| 500 | Higher | Oracle | 0.015 (0.002) | 1.15 | — | — | .960 | .938 | — | 0.214 |
| 500 | Higher | Uncorrected | 0.056 (0.002) | 1.16 | 0.91 | .736 | .838 | .840 | 0.172 | 0.211 |
| 500 | Higher | Croon-DML | 0.019 (0.002) | 1.14 | 0.91 | .894 | .964 | .940 | 0.182 | 0.221 |
| 1000 | Lower | Oracle | 0.007 (0.001) | 1.25 | — | — | .982 | .948 | — | 0.154 |
| 1000 | Lower | Uncorrected | 0.112 (0.001) | 1.22 | 0.92 | .046 | .096 | .192 | 0.118 | 0.150 |
| 1000 | Lower | Croon-DML | 0.021 (0.002) | 1.17 | 0.92 | .886 | .922 | .928 | 0.142 | 0.173 |
| 1000 | Higher | Oracle | 0.010 (0.002) | 1.23 | — | — | .984 | .950 | — | 0.156 |
| 1000 | Higher | Uncorrected | 0.057 (0.001) | 1.24 | 0.94 | .542 | .692 | .730 | 0.121 | 0.154 |
| 1000 | Higher | Croon-DML | 0.015 (0.002) | 1.22 | 0.94 | .912 | .960 | .946 | 0.130 | 0.161 |
| 5000 | Lower | Oracle | 0.002 (0.001) | 1.16 | — | — | .978 | .954 | — | 0.064 |
| 5000 | Lower | Uncorrected | 0.118 (0.001) | 1.16 | 0.98 | .000 | .000 | .000 | 0.053 | 0.061 |
| 5000 | Lower | Croon-DML | 0.008 (0.001) | 1.17 | 1.01 | .932 | .920 | .960 | 0.067 | 0.075 |
| 5000 | Higher | Oracle | 0.002 (0.001) | 1.16 | — | — | .970 | .964 | — | 0.064 |
| 5000 | Higher | Uncorrected | 0.055 (0.001) | 1.14 | 0.96 | .028 | .038 | .066 | 0.054 | 0.063 |
| 5000 | Higher | Croon-DML | 0.005 (0.001) | 1.13 | 0.96 | .928 | .944 | .954 | 0.059 | 0.068 |

*Note.* Abbreviations and computations are as in Table S8.

## S4. Computational details

### *S4.1 LOOCV for the penalty parameter, smoother weights, learners, and software*

In a training fold, the matrix $\tilde{Z}$ of standardized second-degree polynomial terms and the target vector $t$ are centered, and $\tilde{Z}_c = QSP'$ ($S = \text{diag}(s_j)$) is the singular value decomposition of the centered design matrix. The ridge fitted values for penalty parameter $\alpha$ are $\hat{t} = Q\text{diag}\{s_j^2/(s_j^2+\alpha)\}Q't$, and the leverage of training point $i$ is

$$H_{ii} = n^{-1} + \sum_j q_{ij}^2 \frac{s_j^2}{s_j^2+\alpha}, \qquad (26)$$

so the leave-one-out residuals $r_i/(1-H_{ii})$ are computed exactly without refitting (Golub et al., 1979; Hastie et al., 2009). For each target, the $\alpha$ that minimizes the mean square of these residuals is selected from 16 log-spaced candidates between $10^{-1}$ and $10^4$ (25 candidates between $10^{-1}$ and $10^7$ in the empirical illustration and in the extended-range condition of Section S3.1); the intercept is not penalized. When the number of terms exceeds the training sample size, the same quantities are computed from the eigendecomposition of the $n \times n$ kernel matrix. Common-penalty LOOCV selects, for the three conditional means jointly, the single $\alpha$ that minimizes the sum over targets of the mean squared leave-one-out residuals, each divided by the training variance of its target; in the simulations each estimator selects its penalty from the targets it observes ($Y, D, \eta_1$–$\eta_3$ for the oracle; $Y, D, S_1$–$S_3$ for the factor-score estimators), and in the empirical illustration from $Y, D, S_1, S_2$.

The prediction-weight vector at an evaluation point $z$ is

$$w_\alpha(z)' = n^{-1}\mathbf{1}' + (\tilde{z}-\bar{\tilde{z}})'P\text{diag}\left\{\frac{s_j}{s_j^2+\alpha}\right\}Q', \qquad (27)$$

with $\| w_\alpha(z) \|^2 = n^{-1} + \sum_j \{p_j'(\tilde{z}-\bar{\tilde{z}})\}^2 s_j^2/(s_j^2+\alpha)^2$, whose mean over evaluation points is $\bar{\omega}$. If components $j, k$ use penalties $\alpha_j, \alpha_k$, the row-specific matrices of Croon-DML-R (equation (15) of the main text) are

$$(G_i)_{jk} = w_{\alpha_j}(z_i)'w_{\alpha_k}(z_i) = n^{-1} + \sum_l \{p_l'(\tilde{z}_i - \bar{\tilde{z}})\}^2 \frac{s_l^2}{(s_l^2+\alpha_j)(s_l^2+\alpha_k)}, \qquad (28)$$

computed from the same singular value decomposition; their mean over held-out rows is $\bar{G}$ of equation (S8), which reduces to $\bar{\omega}$ when all components share one $\alpha$.

The tree learners of Section S3.1 are histogram-based gradient boosting (Friedman, 2001; 250 iterations, learning rate 0.05, 31 leaves, no L2 penalty) and random forests (Breiman, 2001; 300 trees, minimum leaf size 5, 70% of the variables considered at each split), with fixed random seeds. Analyses were carried out in Python 3.13 with NumPy (Harris et al., 2020), SciPy (Virtanen et al., 2020), and scikit-learn (Pedregosa et al., 2011). The measurement model estimation was cross-checked against Mplus 9.1 (Muthén & Muthén, 2026) on one sample generated under the conditions of the main design ($N = 5000$, lower measurement quality, strong confounding; a separate seed), with factor variances fixed to 1, factor means to 0, and intercepts, loadings, and error variances free: the maximum absolute difference in the intercepts, loadings, error variances and covariances, factor correlations, Bartlett coefficients, and Ω was 0.000699 (diagonal Θ) and 0.000872 (with the two error covariances of Section S2.4), within the rounding of the three-decimal Mplus output. Mplus was also used to cross-check the MNLFA of Section S5.3.

### *S4.2 Rules for numerical failures*

In each fold, the measurement model is optimized once with the normal-theory ML objective and analytic gradients; convergence requires a finite objective and either optimizer success or a maximum absolute gradient of at most $10^{-4}$. Error variances are estimated on the log scale, so negative variances cannot occur; an error variance below 1% of the sample variance of its indicator is recorded as an improper solution. Prespecified error correlations are kept admissible by a hyperbolic tangent transformation, and the positive definiteness of the error covariance and factor correlation matrices and the column rank of the loading matrix are checked after optimization. Nonconvergence or improper solutions of the measurement model, failures in computing the Bartlett scores or the conditional means, a corrected latent block or $\hat{A}$ that is not positive definite, and a condition number of $\hat{A}$ that is not finite or exceeds $10^{10}$ are recorded as estimation failures; nothing is retried or adjusted, and negative variances or boundary solutions are not fixed at zero or small values (F. Chen et al., 2001). A failure of one estimator does not discard the other estimators of the same replication. No estimation failures occurred in the main design or in any supplementary simulation; in the two bootstrap studies, 70 of 42,000 method-level results (0.17%) yielded no interval because the re-estimation success criterion was not met (Sections S3.2 and S9).

## S5. Data, measurement model, and measurement invariance in the empirical illustration

### *S5.1 Data and variables*

Section 6 of the main text uses variables from the 2009 base-year survey and the 2012 first follow-up of the HSLS:09 public-use student file (National Center for Education Statistics, 2018; Duprey et al., 2018; Table S10). Of the 23,503 students, 20,024 have the outcome, 17,790 also the study-time item, 14,351 also the six indicators, prior achievement, and SES, 11,594 also the variables derived from the parent questionnaire, and 9,491 the remaining baseline covariates. The complete-case analysis assumes that the data are missing at random (MAR) given the observed variables; a multiple imputation sensitivity analysis is reported in Section S8.2. Because both focal variables were measured in 2012, the analysis concerns their concurrent association adjusted for baseline covariates.

**Table S10.** Variables in the HSLS:09 illustration

| Role | Wave | Variable | Definition |
|---|---|---|---|
| Outcome | 2012 | Mathematics identity scale score | X2MTHID (standardized score of two items, "sees self as a math person" and "others see self as a math person") |
| Focal predictor | 2012 | Weekly hours of mathematics homework and study | S2MHOMEWRK (time spent on mathematics homework and studying in a typical school week: 0 = none, 1 = less than 30 minutes, 2 = 30 minutes–1 hour, 3 = 1–2 hours, 4 = 2–3 hours, 5 = 4–6 hours, 6 = 7–9 hours, 7 = more than 9 hours), recoded to representative times of 0, 0.25, 0.75, 1.5, 2.5, 5, 8, and 11 hours (11 hours is an assigned value for the open-ended last category); analysis-sample frequencies 601/1,675/2,573/1,906/1,367/969/280/120 |
| Indicators | 2009 | Two mathematics identity items | S1MPERSON1, S1MPERSON2 (four-point; reverse-coded so that higher values indicate stronger identity) |
| | 2009 | Four mathematics self-efficacy items | S1MTESTS (tests), S1MTEXTBOOK (textbook), S1MSKILLS (mastering skills), S1MASSEXCL (assignments) (four-point; reverse-coded) |
| Observed covariates (33) | 2009 | Prior achievement | X1TXMTH (ninth-grade mathematics assessment $\theta$ score) |

| Role | Wave | Variable | Definition |
|---|---|---|---|
| | 2009 | Socioeconomic background | X1SES, X1PAREDU (parents' highest education), X1FAMINCOME (family income category), X1HHNUMBER (household size) |
| | 2009 | Educational expectations | X1STUEDEXPCT, X1PAREDEXPCT (1–10; "don't know" coded with a separate indicator variable and replaced by the median) |
| | 2009 | Time use | S1HRMHOMEWK (mathematics homework), S1HRSHOMEWK (science homework), S1HRACTIVITY (extracurricular activities), S1HRWORK (paid work), S1HRTV, S1HRVIDEO, S1HRONLINE (interval categories for a typical weekday, from 1 = less than 1 hour to 6 = 5 hours or more; treated as ordinal scores) |
| | 2009 | Demographics and school | Female, race/ethnicity (eight categories, seven indicator variables), home language (X1DUALLANG; three categories, two indicator variables), private school (X1CONTROL), locale (X1LOCALE; four categories, three indicator variables), region (X1REGION; four categories, three indicator variables) |

*Note.* Variable names are those of the HSLS:09 public-use file. HSLS:09 = High School Longitudinal Study of 2009; SES = socioeconomic status composite.

### *S5.2 Measurement model and invariance across gender*

Table S11 reports the item distributions and the correlated two-factor measurement model of the main text; all six items met the normality criteria of Curran et al. (1996; |skewness| < 2, kurtosis < 7). Because gender is one of the observed covariates, metric and scalar invariance across gender were examined by multigroup normal-theory ML with mean structures (Table S12). With boys ($n = 4{,}649$) as the reference group (factor variances 1, factor means 0), the factor variances and covariances of girls ($n = 4{,}842$) were free from the metric model onward and their factor means in the scalar model (Meredith, 1993). $\Delta\chi^2$ was statistically significant at both steps, but ΔCFI was −.0003 and −.0042, ΔRMSEA at most .0072, and ΔSRMR at most .005, within the large-sample criteria of F. F. Chen (2007): a CFI decrease of at most .010, supplemented by an RMSEA increase of at most .015 or an SRMR increase of at most .030 for the loadings and .010 for the intercepts (see also Cheung & Rensvold, 2002).

**Table S11.** Item distributions and measurement model in the illustration (correlated two-factor model, $N = 9{,}491$)

| Factor | Item | Mean | SD | Skewness | Kurtosis | Standardized loading | Communality | Error variance |
|---|---|---|---|---|---|---|---|---|
| Identity | S1MPERSON1 | 2.59 | 0.94 | −0.149 | −0.859 | .900 | .809 | .191 |
| Identity | S1MPERSON2 | 2.62 | 0.90 | −0.212 | −0.725 | .805 | .649 | .351 |
| Self-efficacy | S1MTESTS | 3.04 | 0.74 | −0.472 | 0.038 | .840 | .706 | .294 |
| Self-efficacy | S1MTEXTBOOK | 2.78 | 0.81 | −0.245 | −0.427 | .793 | .629 | .371 |
| Self-efficacy | S1MSKILLS | 3.05 | 0.70 | −0.459 | 0.239 | .837 | .701 | .299 |

| Factor | Item | Mean | SD | Skewness | Kurtosis | Standardized loading | Communality | Error variance |
|---|---|---|---|---|---|---|---|---|
| Self-efficacy | S1MASSEXCL | 3.14 | 0.69 | −0.571 | 0.482 | .851 | .724 | .276 |

*Note.* Four-point items, reverse-coded so that higher values indicate stronger identity or efficacy. SD = standard deviation; skewness and (excess) kurtosis are sample moments (Curran et al., 1996). Factor variances are fixed to 1; the factor correlation is .667. Fit: $\chi^2(8) = 259.8$, TLI = .986, CFI = .993, RMSEA = .058, SRMR = .014. Standardized loadings are the loadings estimated with unit factor variances divided by the sample standard deviation of the indicator; communality = squared standardized loading; error variance = 1 − communality. Composite reliability (Raykov, 1997) is .845 (identity) and .897 (self-efficacy), Cronbach's $\alpha$ .840 and .897, and the average variance extracted (Fornell & Larcker, 1981; mean of the squared unrounded standardized loadings) .729 and .690, both above the squared factor correlation of .444. The identity factor has two indicators and is identified through its correlation with self-efficacy. TLI = Tucker–Lewis index; CFI = comparative fit index; RMSEA = root mean square error of approximation; SRMR = standardized root mean square residual.

**Table S12.** Tests of measurement invariance across gender (boys $n = 4{,}649$, girls $n = 4{,}842$)

| Model | $\chi^2$ | $df$ | TLI | CFI | RMSEA | SRMR | $\Delta\chi^2$ ($\Delta df$) | ΔCFI | ΔRMSEA | ΔSRMR |
|---|---|---|---|---|---|---|---|---|---|---|
| Configural invariance | 262.4 | 16 | .986 | .993 | .057 | .012 | — | — | — | — |
| Metric invariance (equal loadings) | 275.9 | 20 | .989 | .992 | .052 | .013 | 13.5 (4) | −.0003 | −.0050 | .001 |
| Scalar invariance (equal loadings and intercepts) | 422.0 | 24 | .985 | .988 | .059 | .018 | 146.0 (4) | −.0042 | .0072 | .005 |

*Note.* Multigroup normal-theory maximum likelihood with mean structure; the reference group is boys (factor variances 1, factor means 0). $\chi^2$ is the sum over groups of the fit function multiplied by $(n_g - 1)$; RMSEA follows the multigroup definition; SRMR is the root mean square of the standardized covariance and mean residuals averaged over the two groups. Δ = difference from the preceding model. In the scalar model, the factor means of girls were −0.18 (identity) and −0.23 (self-efficacy). TLI = Tucker–Lewis index; CFI = comparative fit index; RMSEA = root mean square error of approximation; SRMR = standardized root mean square residual.

### *S5.3 Invariance with respect to a continuous covariate (MNLFA)*

The first condition of A2 requires invariance of the loadings and intercepts with respect to all observed covariates. Moderated nonlinear factor analysis (Bauer, 2017) was applied to the complete-case sample with the strongest observed covariate, standardized prior mathematics achievement, as the moderator. The impact model specifies the factor means as linear and the factor variances as exponential-linear functions of the moderator, with invariant loadings and intercepts; each item-level DIF model additionally lets the loading and intercept of one item depend linearly on the moderator, with the remaining items as anchors, and is compared with the impact model by a likelihood ratio test with two degrees of freedom. The models were fitted by normal-theory ML; refitting the impact model and the DIF model for S1MPERSON1 in Mplus 9.1 reproduced the log-likelihoods and all estimates to three decimals.

In the impact model, the moderation coefficients of the factor means were 0.507 (identity) and 0.359 (self-efficacy; SEs 0.013 and 0.011), and those of the log factor variances −0.166 and −0.149 (SEs 0.018 and 0.015). The $\Delta\chi^2(2)$ statistics were 128.1 (S1MPERSON1), 124.0 (S1MPERSON2), 5.0 (S1MTESTS), 34.7 (S1MTEXTBOOK), 10.0 (S1MSKILLS), and 22.1 (S1MASSEXCL), against a critical value of 5.99. Effect sizes are the loading moderation coefficient divided by the baseline loading (proportional change per 1 SD of the moderator) and the intercept moderation coefficient divided by the item SD (shift per 1 SD, in item SD units). For the self-efficacy items, loading changes (−0.031 to 0.049) and intercept shifts (−0.015 to 0.022) were small. For the identity items, loading changes were small (0.002–0.015), but the intercept shifts were −0.123 (S1MPERSON1, anchored on S1MPERSON2) and 0.102 (S1MPERSON2, anchored on S1MPERSON1): uniform DIF, with the two intercepts moving in opposite directions (with two indicators, only relative DIF is identified). A model with error variances that depend exponential-linearly on the moderator fitted better than the impact model ($\Delta\chi^2(6) = 399.1$; slopes of the log error variances 0.030 and −0.052 for the identity items and −0.105 to −0.212 for the self-efficacy items), so the invariance of the error variances in condition C2 (needed only for the approximation of equation (14)) does not hold strictly; this has no bearing on identification or consistency.

DIF can affect the correction in two ways. First, because the moderator is one of the observed covariates $Z$, the intercept DIF term is a function of $Z$, is absorbed into the conditional mean of the factor scores, and does not remain in the residualized scores. Second, a measurement model that ignores DIF may change $\widehat{\Theta}$ and $\widehat{\Omega}$. To check the second route, the cross-fitted residuals, folds, and Bartlett weights of Table 6 of the main text were kept and only the subtracted error covariance was recomputed as $\widehat{B}_k\widehat{\Theta}_{\mathrm{DIF}}\widehat{B}_k'$ from a model that allows DIF (moderated loadings and intercepts for one identity item and three self-efficacy items). The diagonal element of $\widehat{\Omega}$ for identity changed from 0.164 to 0.160–0.161, that for self-efficacy remained 0.110, and the factor-score error correction changed from 0.00104 to 0.00101–0.00102, leaving the conclusions of the main text unchanged. With the socioeconomic status composite as the moderator, the intercept shifts were 0.05–0.06 item SD per 1 SD for the identity items and at most 0.03 for the self-efficacy items, and $\widehat{\Omega}$ changed by less than 2%.

## S6. Nonlinear latent conditional mean

Because $q_0$ is linear in the main design and in Section S3.1, nonlinearity in $q_0$ itself was examined separately. With $Z \sim N(0, I_{20})$, the residual latent covariance, the measurement conditions, strong confounding, and $\theta_0 = 0.30$ of the main design, $q_0$ was changed to

$$q_{0j}(Z) = 0.25\left\{Z_{2j-1} + \frac{Z_{2j}^2 - 1}{\sqrt{2}}\right\}, \qquad j = 1,2,3, \qquad (29)$$

so that each component has mean 0 and variance 0.125 and the unconditional latent variance of 1 and the latent correlations of .30 are unchanged. Because the latent variables are then nonnormal, this check also evaluates the normal-theory ML fit of a correctly specified linear measurement model. In four conditions ($N = 1000, 5000$ × lower/higher measurement quality; 200 replications each), a setting in which all three conditional means are learned by quadratic polynomial ridge regression was compared, on the same data and folds, with a setting in which only the $q_0$ learner is replaced by linear ridge regression, which cannot represent $Z_{2j}^2$ (Table

S13). With the quadratic learner, the bias of Croon-DML at $N = 5000$ was 0.004–0.007 against 0.055–0.119 for uncorrected DML; with the misspecified $q_0$ learner, Croon-DML and the oracle were both biased by about 0.038–0.039. Factor-score error correction cannot compensate for a misspecified conditional mean.

**Table S13.** Bias and RMSE depending on whether the learner represents the nonlinear $q_0$ (strong confounding, $\theta_0 = 0.30$, 200 paired replications per condition)

| $N$ | Measurement quality | $q_0$ learner | Oracle bias | Uncorrected bias | Croon bias | Oracle RMSE | Uncorrected RMSE | Croon RMSE |
|---|---|---|---|---|---|---|---|---|
| 1000 | Lower | Quadratic polynomial ridge regression | 0.012 | 0.120 | 0.022 | 0.035 | 0.124 | 0.044 |
| 1000 | Lower | Linear ridge regression (misspecified) | 0.060 | 0.144 | 0.059 | 0.068 | 0.147 | 0.069 |
| 1000 | Higher | Quadratic polynomial ridge regression | 0.009 | 0.057 | 0.014 | 0.035 | 0.066 | 0.037 |
| 1000 | Higher | Linear ridge regression (misspecified) | 0.058 | 0.095 | 0.058 | 0.066 | 0.100 | 0.066 |
| 5000 | Lower | Quadratic polynomial ridge regression | 0.003 | 0.119 | 0.007 | 0.015 | 0.120 | 0.019 |
| 5000 | Lower | Linear ridge regression (misspecified) | 0.039 | 0.138 | 0.038 | 0.042 | 0.138 | 0.042 |
| 5000 | Higher | Quadratic polynomial ridge regression | 0.001 | 0.055 | 0.004 | 0.015 | 0.057 | 0.016 |
| 5000 | Higher | Linear ridge regression (misspecified) | 0.038 | 0.083 | 0.038 | 0.041 | 0.084 | 0.041 |

*Note.* Oracle = oracle latent-variable DML; uncorrected = uncorrected factor-score DML; Croon = Croon-DML; RMSE = root mean square error; DML = double/debiased machine learning. The learners for the outcome and the focal predictor are quadratic polynomial ridge regression in all settings. Monte Carlo standard errors of bias are 0.0010–0.0028; there were no estimation failures.

**S7. Croon-DML-R and the common-penalty learner**

The paired studies reported in Sections 5.1 and 5.3 of the main text use the seeds and folds of the 12 conditions of the main design, so they are paired replication by replication with the main design and with Table S6. Table S14 adds Croon-DML-R to ridge regression with $\alpha = 1$ (the 250 replications of Table S6). Under target-specific LOOCV (1,000 replications per condition), the four estimators are reported in Table 5 and Figure 2 of the main text; their oracle, uncorrected, and Croon-DML estimates equal those of the main design to within $10^{-8}$ in every replication. Table S15 reports the four estimators under the common-penalty LOOCV learner of Section S4.1 (1,000 replications per condition), which supplies the common-penalty columns of Table 5. None of these studies had estimation failures.

**Table S14.** Performance of Croon-DML and Croon-DML-R with a fixed penalty ($\alpha = 1$; 12 conditions of the main design, 250 paired replications per condition)

| $N$ | Measurement quality | Confounding | Oracle bias | Uncorrected bias | Croon-DML bias | Croon-DML-R bias | Uncorrected RMSE | Croon-DML RMSE | Croon-DML-R RMSE | Croon − oracle | Croon-R − oracle | $\bar{\omega}$ | Equation (14) approximation |
|---|---|---|---|---|---|---|---|---|---|---|---|---|---|
| 500 | Lower | Moderate | −0.005 | 0.062 | 0.043 | −0.005 | 0.102 | 0.093 | 0.093 | 0.047 | 0.000 | 1.64 | 0.048 |
| 500 | Lower | Strong | −0.004 | 0.121 | 0.090 | −0.001 | 0.143 | 0.120 | 0.102 | 0.094 | 0.003 | 1.64 | 0.087 |
| 500 | Higher | Moderate | −0.001 | 0.028 | 0.018 | −0.001 | 0.081 | 0.079 | 0.079 | 0.019 | 0.000 | 1.64 | 0.019 |
| 500 | Higher | Strong | −0.002 | 0.054 | 0.036 | −0.001 | 0.096 | 0.090 | 0.088 | 0.038 | 0.001 | 1.64 | 0.036 |
| 1000 | Lower | Moderate | 0.002 | 0.070 | 0.029 | 0.001 | 0.081 | 0.052 | 0.046 | 0.027 | −0.001 | 0.45 | 0.027 |
| 1000 | Lower | Strong | −0.003 | 0.116 | 0.049 | −0.003 | 0.122 | 0.066 | 0.050 | 0.052 | 0.000 | 0.45 | 0.051 |
| 1000 | Higher | Moderate | −0.000 | 0.028 | 0.009 | −0.001 | 0.052 | 0.046 | 0.045 | 0.010 | −0.000 | 0.45 | 0.010 |
| 1000 | Higher | Strong | 0.000 | 0.055 | 0.021 | 0.001 | 0.071 | 0.051 | 0.049 | 0.021 | 0.001 | 0.45 | 0.019 |
| 5000 | Lower | Moderate | −0.001 | 0.066 | 0.005 | −0.001 | 0.068 | 0.017 | 0.017 | 0.006 | 0.000 | 0.06 | 0.006 |
| 5000 | Lower | Strong | −0.000 | 0.118 | 0.012 | −0.000 | 0.119 | 0.020 | 0.017 | 0.012 | 0.000 | 0.06 | 0.012 |
| 5000 | Higher | Moderate | −0.002 | 0.027 | 0.001 | −0.001 | 0.031 | 0.016 | 0.016 | 0.002 | 0.000 | 0.06 | 0.002 |
| 5000 | Higher | Strong | 0.000 | 0.055 | 0.005 | 0.001 | 0.057 | 0.017 | 0.017 | 0.005 | 0.001 | 0.06 | 0.004 |

*Note.* Bias = mean estimate − $\theta_0$; RMSE = root mean square error; oracle = oracle latent-variable DML; uncorrected = uncorrected factor-score DML; Croon-R = Croon-DML-R (residualization-inflation correction); DML = double/debiased machine learning. The difference columns are means of the within-replication differences between the two estimates. $\bar{\omega}$ = mean over held-out rows of $\| w_i \|^2$ (the inflation factor minus 1); Equation (14) approximation = approximate residual bias of Croon-DML obtained by substituting $\bar{\omega}$ into equation (14) of the main text.

**Table S15.** Performance of the four estimators with the common-penalty LOOCV learner (12 conditions of the main design, 1,000 paired replications per condition)

| $N$ | Measurement quality | Confounding | Oracle bias | Uncorrected bias | Croon-DML bias | Croon-DML-R bias | Uncorrected RMSE | Croon-DML RMSE | Croon-DML-R RMSE | Croon − oracle | Croon-R − oracle |
|---|---|---|---|---|---|---|---|---|---|---|---|
| 500 | Lower | Moderate | 0.010 | 0.062 | 0.014 | 0.011 | 0.076 | 0.049 | 0.049 | 0.004 | 0.001 |
| 500 | Lower | Strong | 0.009 | 0.099 | 0.018 | 0.012 | 0.107 | 0.052 | 0.051 | 0.009 | 0.003 |
| 500 | Higher | Moderate | 0.009 | 0.030 | 0.011 | 0.010 | 0.054 | 0.047 | 0.047 | 0.002 | 0.001 |
| 500 | Higher | Strong | 0.011 | 0.049 | 0.014 | 0.012 | 0.066 | 0.049 | 0.048 | 0.003 | 0.001 |
| 1000 | Lower | Moderate | 0.009 | 0.064 | 0.013 | 0.009 | 0.071 | 0.036 | 0.035 | 0.004 | 0.000 |
| 1000 | Lower | Strong | 0.006 | 0.104 | 0.017 | 0.010 | 0.108 | 0.039 | 0.037 | 0.011 | 0.004 |

| | | | | | | | | | | |
|---|---|---|---|---|---|---|---|---|---|---|
| 1000 | Higher | Moderate | 0.008 | 0.031 | 0.010 | 0.009 | 0.044 | 0.033 | 0.032 | 0.002 | 0.001 |
| 1000 | Higher | Strong | 0.007 | 0.049 | 0.010 | 0.007 | 0.058 | 0.033 | 0.033 | 0.003 | 0.001 |
| 5000 | Lower | Moderate | 0.001 | 0.067 | 0.005 | 0.002 | 0.069 | 0.017 | 0.016 | 0.004 | 0.000 |
| 5000 | Lower | Strong | 0.001 | 0.116 | 0.008 | 0.001 | 0.117 | 0.019 | 0.017 | 0.007 | 0.000 |
| 5000 | Higher | Moderate | 0.000 | 0.028 | 0.002 | 0.001 | 0.032 | 0.016 | 0.016 | 0.002 | 0.000 |
| 5000 | Higher | Strong | 0.001 | 0.053 | 0.004 | 0.002 | 0.055 | 0.016 | 0.015 | 0.003 | 0.000 |

*Note.* Abbreviations are as in Table S14. The common penalty is the single $\alpha$ that minimizes the sum of the standardized target-specific LOOCV errors and is used for all three conditional means (Section S4.1). LOOCV = leave-one-out cross-validation.

## S8. Bootstrap, multiple imputation, and outcome sensitivity in the empirical illustration

### *S8.1 Paired bootstrap for Table 6*

The specification without prior mathematics achievement uses 32 observed covariates (560 second-degree polynomial terms) on the same sample as the specification with all 33 covariates (594 terms). The bootstrap resamples students; in each resample, all copies of an original observation are placed in the same fold, and the measurement model, $\hat{B}, \hat{\Omega}$, the conditional means, and the structural coefficients are re-estimated. The conditional means are learned with the recommended learner, quadratic polynomial ridge regression with one common LOOCV penalty for the four targets $Y, D, S_1, S_2$ selected from 25 candidates between $10^{-1}$ and $10^{7}$; the selected penalty was $10^{4}$ in every fold of both specifications, well inside the candidate range. The penalty selected in the original sample is held fixed in all resample folds (the fixed aggregated rule of Section S3.2; with one common penalty, its per-target geometric mean is the same value for every target), so the intervals of Table 6 of the main text come from the procedure whose coverage Table S18 evaluates.

The intervals for the latent-covariate adjustment and the factor-score error correction are BC intervals for the difference between the two estimates in the same resamples; both differences are defined relative to the recommended estimator Croon-DML-R, whose mean inflation factor was 1.007 (with prior achievement) and 1.006 (without) and whose estimates differed from those of Croon-DML by less than 0.0001. With target-specific LOOCV penalties (4,642–10,000), the estimates of all four estimators differed from those reported by at most 0.0001, and the stacked sandwich SE of Section S1.4 equalled the conditional sandwich SE to four decimals (0.0043 with and 0.0044 without prior achievement) for the uncorrected estimator, Croon-DML, and Croon-DML-R, so at $N = 9{,}491$ the estimation of the measurement parameters contributes negligibly to the uncertainty; the bootstrap SEs (0.0044–0.0046) exceed both by 2–5%, for the reason given in Section S3.2.

### *S8.2 Multiple imputation sensitivity analysis*

The complete-case analysis uses 9,491 of the 23,503 students. Under the MAR assumption, the analysis was repeated with multiple imputation (Rubin, 1987; Enders, 2022) for the 17,790 students whose 2012 outcome and study time were observed; the baseline covariates, the six indicators, and the parent-questionnaire variables were imputed, whereas 2012 variables were not. The imputation model used chained equations (van Buuren, 2018) with eight cycles of Bayesian ridge regressions with posterior draws and, as auxiliary variables, 2012 mathematics achievement, 2012 SES, the baseline mathematics self-efficacy and interest scale scores, and the 2012 self-efficacy scale score. Binary variables were rounded, exactly one indicator was activated within each block of categorical indicator variables, ordinal variables were rounded and truncated to their range, and the four-point indicators were truncated to [1, 4] and kept continuous. In each of the $m = 10$ imputed data sets, the four estimators were computed with the seed, folds, learner (common-penalty LOOCV), and measurement model of Section 6 of the main text, and the results were combined by Rubin's rules with the conditional sandwich SE as the within-imputation variance (Table S16).

**Table S16.** Complete-case and pooled multiple imputation estimates in HSLS:09

| Specification | Estimator | Complete-case estimate | Pooled MI estimate (SE) | Between-imputation SD | FMI |
|---|---|---|---|---|---|
| With prior achievement (33) | DML without latent covariates | 0.0275 | 0.0322 | 0.0006 | — |
| With prior achievement (33) | Uncorrected factor-score DML | 0.0215 | 0.0260 (0.0032) | 0.0006 | .0347 |
| With prior achievement (33) | Croon-DML | 0.0205 | 0.0246 (0.0032) | 0.0006 | .0403 |
| With prior achievement (33) | Croon-DML-R | 0.0205 | 0.0246 (0.0032) | 0.0006 | .0404 |
| Without prior achievement (32) | DML without latent covariates | 0.0462 | 0.0482 | 0.0005 | — |
| Without prior achievement (32) | Uncorrected factor-score DML | 0.0294 | 0.0329 (0.0033) | 0.0005 | .0313 |
| Without prior achievement (32) | Croon-DML | 0.0268 | 0.0301 (0.0033) | 0.0006 | .0409 |
| Without prior achievement (32) | Croon-DML-R | 0.0268 | 0.0301 (0.0033) | 0.0006 | .0410 |

*Note.* Multiple imputation sample $N = 17{,}790$ (2012 outcome and study time observed), $m = 10$; complete cases $N = 9{,}491$. The SE of the pooled estimate follows Rubin's (1987) rules with the conditional sandwich standard error as the within-imputation variance; for DML without latent covariates, only the point estimates were pooled. MI = multiple imputation; SD = standard deviation; FMI = fraction of missing information; DML = double/debiased machine learning; HSLS:09 = High School Longitudinal Study of 2009.

### *S8.3 Outcome sensitivity analysis*

The outcome of Section 6 of the main text is the scale score of two 2012 mathematics identity items, and the indicators of the latent covariate 2009 mathematics identity are the 2009 responses to the same two items. If item-specific components are stable over time, the indicator errors and the structural error of the outcome are correlated, violating $E(\varepsilon e \mid \eta, Z) = 0$; by equations (S10)–(S11) with $c_v = 0$, the effect on the structural coefficient is $-\gamma_0' c_e/\mathrm{Var}(v)$, proportional to $\gamma_0$. Because the size of the violation cannot be estimated directly, the outcome was replaced by the 2012 mathematics achievement $\theta$ score, which shares no items with the indicators, and the procedure of Section 6 was repeated (same covariate specifications, measurement model, learner, seed, and 300-resample bootstrap; complete-case $N = 9{,}636$ because of the different missingness pattern; Table S17). The ordering of the estimators and the direction of the correction were unchanged. The adjustment (0.0019 with and 0.0111 without prior achievement) and the correction (0.0003 and 0.0015) were smaller than for the identity outcome because 2012 achievement correlates .75 with prior achievement, so $\hat{\beta}_C$ is small (0.14 for identity vs. 0.56 in the main text); the correction remained 15% and 13% of the adjustment, as in the main text. In standardized terms, the Croon-DML-R coefficient is .059 SD (with) or .107 SD (without prior achievement) of mathematics achievement per 1 SD of study time, and Croon-DML-R agreed with Croon-DML to within 0.0001 (mean inflation factors 1.010 and 1.007; common penalties 6,310 and 8,577).

**Table S17.** Sensitivity analysis in HSLS:09 with 2012 mathematics achievement as the outcome, by estimator and observed-covariate specification

| Estimator | With prior achievement (33) | Without prior achievement (32) |
|---|---|---|
| DML without latent covariates | 0.0339 (0.0039) [0.0270, 0.0422] | 0.0694 (0.0055) [0.0596, 0.0801] |
| Uncorrected factor-score DML | 0.0324 (0.0040) [0.0257, 0.0406] | 0.0597 (0.0053) [0.0499, 0.0690] |
| Croon-DML | 0.0321 (0.0040) [0.0255, 0.0404] | 0.0583 (0.0054) [0.0483, 0.0683] |
| Croon-DML-R | 0.0321 (0.0040) [0.0255, 0.0404] | 0.0583 (0.0054) [0.0483, 0.0683] |
| Latent-covariate adjustment (no latent − Croon-DML-R) | 0.0019 [0.0003, 0.0034] | 0.0111 [0.0074, 0.0153] |
| Factor-score error correction (uncorrected − Croon-DML-R) | 0.0003 [−0.0000, 0.0006] | 0.0015 [0.0008, 0.0023] |

*Note.* $N = 9{,}636$ (complete cases with 2012 mathematics achievement observed). Cells show the point estimate (bootstrap SE) [bias-corrected 95% interval]; for the adjustment and the correction, the interval is the bias-corrected 95% interval of the difference between the two estimates in the same resamples (300 resamples, no failures; rules as in Table 6 of the main text). The residualized $\hat{\gamma}$ is $(0.057, 0.007)'$ with and $(0.131, 0.014)'$ without prior achievement, and the residualized reliabilities are (.83, .89) and (.85, .89). DML = double/debiased machine learning; HSLS:09 = High School Longitudinal Study of 2009; SE = standard error.

**S9. Intervals for the recommended combination**

The interval performance of the combination recommended in Section 7 of the main text (common-penalty LOOCV, Croon-DML-R, and a BC bootstrap that also refits the measurement model) was evaluated for the four estimators in the design of Section S3.2 (Table S18); the only difference from Tables S8–S9 is the common-penalty learner. In the same 14 samples as in Section S3.2 ($N = 500$, lower measurement quality), the re-estimation success rates of uncorrected DML, Croon-DML, and Croon-DML-R fell below .90, so the corresponding cells are based on 493 replications. With the common penalty, the bias of Croon-DML was smaller than in Tables S8–S9 (for example, 0.016 instead of 0.024 at $N = 1000$, lower quality, $\theta_0 = 0$), and Croon-DML-R roughly halved it again, to within 0.001 of the oracle at $N = 5000$. The coverage of Croon-DML-R was .934–.962 (Wald), .950–.970 (percentile), and .930–.962 (BC), and its Type I error rates at $\theta_0 = 0$ were .038–.064 (Wald) and .038–.062 (BC), within twice the MCSE of coverage of the nominal .05. Croon-DML with the same learner had coverage of .896–.958 (Wald) and .920–.960 (BC), lower than Croon-DML-R at $N = 5000$ with lower quality, where residual bias remains. The ratio of bootstrap SE to SD was 1.04–1.18, above 1 for the reason given in Section S3.2, and the ratio of conditional sandwich SE to SD was 0.95–1.06.

**Table S18.** Performance of approximate intervals for the recommended combination (common-penalty LOOCV and Croon-DML-R; strong confounding, 500 outer replications × 300 bootstrap resamples)

| $N$ | Measurement quality | $\theta_0$ | Method | Bias (MCSE) | Boot SE/SD | Sandwich SE/SD | Wald coverage | Percentile coverage | BC coverage |
|---|---|---|---|---|---|---|---|---|---|
| 500 | Lower | 0 | Oracle | 0.009 (0.002) | 1.18 | — | — | .978 | .952 |
| 500 | Lower | 0 | Uncorrected | 0.100 (0.002) | 1.15 | 1.06 | .351 | .408 | .410 |
| 500 | Lower | 0 | Croon-DML | 0.019 (0.002) | 1.16 | 1.06 | .949 | .961 | .955 |
| 500 | Lower | 0 | Croon-DML-R | 0.013 (0.002) | 1.16 | 1.06 | .957 | .970 | .955 |
| 500 | Higher | 0 | Oracle | 0.012 (0.002) | 1.17 | — | — | .982 | .958 |
| 500 | Higher | 0 | Uncorrected | 0.050 (0.002) | 1.16 | 1.06 | .816 | .874 | .828 |
| 500 | Higher | 0 | Croon-DML | 0.015 (0.002) | 1.17 | 1.06 | .958 | .966 | .960 |
| 500 | Higher | 0 | Croon-DML-R | 0.013 (0.002) | 1.17 | 1.06 | .962 | .970 | .962 |
| 1000 | Lower | 0 | Oracle | 0.007 (0.001) | 1.16 | — | — | .978 | .958 |
| 1000 | Lower | 0 | Uncorrected | 0.104 (0.001) | 1.17 | 1.05 | .068 | .100 | .114 |
| 1000 | Lower | 0 | Croon-DML | 0.016 (0.002) | 1.14 | 1.02 | .934 | .942 | .942 |
| 1000 | Lower | 0 | Croon-DML-R | 0.008 (0.002) | 1.13 | 1.02 | .944 | .966 | .948 |
| 1000 | Higher | 0 | Oracle | 0.009 (0.001) | 1.13 | — | — | .966 | .938 |
| 1000 | Higher | 0 | Uncorrected | 0.051 (0.001) | 1.13 | 1.00 | .614 | .700 | .676 |
| 1000 | Higher | 0 | Croon-DML | 0.012 (0.001) | 1.11 | 0.99 | .930 | .950 | .940 |
| 1000 | Higher | 0 | Croon-DML-R | 0.010 (0.001) | 1.11 | 0.99 | .942 | .958 | .944 |
| 5000 | Lower | 0 | Oracle | 0.002 (0.001) | 1.04 | — | — | .954 | .944 |
| 5000 | Lower | 0 | Uncorrected | 0.117 (0.001) | 1.06 | 0.95 | .000 | .000 | .000 |

| $N$ | Measurement quality | $\theta_0$ | Method | Bias (MCSE) | Boot SE/SD | Sandwich SE/SD | Wald coverage | Percentile coverage | BC coverage |
|---|---|---|---|---|---|---|---|---|---|
| 5000 | Lower | 0 | Croon-DML | 0.009 (0.001) | 1.06 | 0.95 | .896 | .886 | .934 |
| 5000 | Lower | 0 | Croon-DML-R | 0.002 (0.001) | 1.05 | 0.95 | .936 | .950 | .938 |
| 5000 | Higher | 0 | Oracle | 0.002 (0.001) | 1.14 | — | — | .974 | .956 |
| 5000 | Higher | 0 | Uncorrected | 0.054 (0.001) | 1.15 | 1.02 | .034 | .038 | .048 |
| 5000 | Higher | 0 | Croon-DML | 0.004 (0.001) | 1.13 | 1.01 | .944 | .964 | .946 |
| 5000 | Higher | 0 | Croon-DML-R | 0.002 (0.001) | 1.13 | 1.00 | .950 | .962 | .944 |
| 500 | Lower | 0.30 | Oracle | 0.009 (0.002) | 1.16 | — | — | .974 | .946 |
| 500 | Lower | 0.30 | Uncorrected | 0.100 (0.002) | 1.16 | 1.06 | .337 | .375 | .402 |
| 500 | Lower | 0.30 | Croon-DML | 0.019 (0.002) | 1.14 | 1.04 | .919 | .943 | .931 |
| 500 | Lower | 0.30 | Croon-DML-R | 0.014 (0.002) | 1.13 | 1.03 | .941 | .951 | .935 |
| 500 | Higher | 0.30 | Oracle | 0.012 (0.002) | 1.09 | — | — | .956 | .928 |
| 500 | Higher | 0.30 | Uncorrected | 0.050 (0.002) | 1.08 | 0.98 | .804 | .838 | .816 |
| 500 | Higher | 0.30 | Croon-DML | 0.015 (0.002) | 1.07 | 0.98 | .932 | .952 | .920 |
| 500 | Higher | 0.30 | Croon-DML-R | 0.013 (0.002) | 1.07 | 0.98 | .934 | .956 | .930 |
| 1000 | Lower | 0.30 | Oracle | 0.005 (0.001) | 1.18 | — | — | .982 | .964 |
| 1000 | Lower | 0.30 | Uncorrected | 0.103 (0.001) | 1.15 | 1.02 | .060 | .104 | .134 |
| 1000 | Lower | 0.30 | Croon-DML | 0.016 (0.002) | 1.13 | 1.01 | .926 | .940 | .936 |
| 1000 | Lower | 0.30 | Croon-DML-R | 0.009 (0.002) | 1.13 | 1.01 | .942 | .962 | .940 |
| 1000 | Higher | 0.30 | Oracle | 0.008 (0.001) | 1.17 | — | — | .966 | .958 |
| 1000 | Higher | 0.30 | Uncorrected | 0.050 (0.001) | 1.17 | 1.04 | .642 | .716 | .676 |
| 1000 | Higher | 0.30 | Croon-DML | 0.011 (0.001) | 1.15 | 1.03 | .950 | .966 | .958 |

| $N$ | Measurement quality | $\theta_0$ | Method | Bias (MCSE) | Boot SE/SD | Sandwich SE/SD | Wald coverage | Percentile coverage | BC coverage |
|---|---|---|---|---|---|---|---|---|---|
| 1000 | Higher | 0.30 | Croon-DML-R | 0.008 (0.001) | 1.15 | 1.03 | .954 | .966 | .956 |
| 5000 | Lower | 0.30 | Oracle | 0.001 (0.001) | 1.15 | — | — | .974 | .954 |
| 5000 | Lower | 0.30 | Uncorrected | 0.116 (0.001) | 1.14 | 1.01 | .000 | .000 | .000 |
| 5000 | Lower | 0.30 | Croon-DML | 0.009 (0.001) | 1.16 | 1.03 | .936 | .918 | .948 |
| 5000 | Lower | 0.30 | Croon-DML-R | 0.002 (0.001) | 1.16 | 1.03 | .954 | .958 | .954 |
| 5000 | Higher | 0.30 | Oracle | 0.002 (0.001) | 1.13 | — | — | .972 | .950 |
| 5000 | Higher | 0.30 | Uncorrected | 0.054 (0.001) | 1.12 | 0.99 | .030 | .050 | .042 |
| 5000 | Higher | 0.30 | Croon-DML | 0.005 (0.001) | 1.11 | 0.99 | .938 | .952 | .946 |
| 5000 | Higher | 0.30 | Croon-DML-R | 0.002 (0.001) | 1.11 | 0.99 | .940 | .960 | .954 |

*Note.* Oracle = oracle latent-variable DML; uncorrected = uncorrected factor-score DML; Croon-DML-R = residualization-inflation correction; MCSE = Monte Carlo standard error; Boot SE/SD = mean bootstrap standard error divided by the empirical standard deviation; Sandwich SE/SD = mean conditional sandwich standard error divided by the empirical standard deviation (not applicable to the oracle); BC = bias-corrected interval; LOOCV = leave-one-out cross-validation; DML = double/debiased machine learning. The learner is common-penalty LOOCV quadratic polynomial ridge regression for all four estimators, with the penalty and fold rules of Section S3.2. Coverage is the proportion of intervals containing $\theta_0$; at $\theta_0 = 0$, the Type I error rate is 1 − coverage. With 500 outer replications, the MCSE of coverage is about 0.010. For uncorrected DML, Croon-DML, and Croon-DML-R at $N = 500$ with lower measurement quality, results are based on the 493 of 500 replications without interval estimation failures.

The stacked sandwich SE of Section S1.4 was evaluated on the same 6,000 outer samples without any bootstrap (Table S19); the recomputed point estimates reproduce those of Table S18 to within $5 \times 10^{-8}$. For Croon-DML-R the ratio of the mean stacked SE to the empirical SD was 0.95–1.07 (conditional sandwich SE: 0.95–1.06; bootstrap SE: 1.05–1.17), the measurement parameters contributed at most 2.0% of the stacked variance, and Wald intervals from the stacked SE covered $\theta_0$ in .934–.962 of the samples (conditional SE: .934–.962; BC bootstrap: .930–.962), with Type I error rates at $\theta_0 = 0$ of .038–.062. The stacked SE therefore gives intervals equivalent to the conditional Wald and the BC bootstrap intervals at a small fraction of the bootstrap's cost; for Croon-DML, whose residual bias is larger, Wald coverage was .898–.958 with the stacked SE, and for uncorrected DML .000–.816, as the bias, not the SE, is at fault.

**Table S19.** Conditional and stacked sandwich standard errors for the recommended combination on the samples of Table S18 (common-penalty LOOCV; strong confounding, 500 outer replications per condition)

| $N$ | Measurement quality | $\theta_0$ | Method | Bias (MCSE) | Conditional SE/SD | Stacked SE/SD | Boot SE/SD | Wald coverage (conditional) | Wald coverage (stacked) | BC coverage |
|---|---|---|---|---|---|---|---|---|---|---|
| 500 | Lower | 0 | Uncorrected | 0.101 (0.002) | 1.06 | 1.06 | 1.15 | .348 | .348 | .410 |
| 500 | Lower | 0 | Croon-DML | 0.019 (0.002) | 1.06 | 1.07 | 1.16 | .948 | .952 | .955 |
| 500 | Lower | 0 | Croon-DML-R | 0.013 (0.002) | 1.06 | 1.07 | 1.16 | .956 | .956 | .955 |
| 500 | Lower | 0.30 | Uncorrected | 0.100 (0.002) | 1.06 | 1.06 | 1.15 | .338 | .336 | .402 |
| 500 | Lower | 0.30 | Croon-DML | 0.019 (0.002) | 1.03 | 1.04 | 1.14 | .920 | .928 | .931 |
| 500 | Lower | 0.30 | Croon-DML-R | 0.014 (0.002) | 1.03 | 1.04 | 1.13 | .942 | .942 | .935 |
| 500 | Higher | 0 | Uncorrected | 0.050 (0.002) | 1.06 | 1.06 | 1.16 | .816 | .816 | .828 |
| 500 | Higher | 0 | Croon-DML | 0.015 (0.002) | 1.06 | 1.06 | 1.17 | .958 | .958 | .960 |
| 500 | Higher | 0 | Croon-DML-R | 0.013 (0.002) | 1.06 | 1.06 | 1.17 | .962 | .962 | .962 |
| 500 | Higher | 0.30 | Uncorrected | 0.050 (0.002) | 0.98 | 0.98 | 1.08 | .804 | .804 | .816 |
| 500 | Higher | 0.30 | Croon-DML | 0.015 (0.002) | 0.98 | 0.98 | 1.07 | .932 | .932 | .920 |
| 500 | Higher | 0.30 | Croon-DML-R | 0.013 (0.002) | 0.98 | 0.98 | 1.07 | .934 | .934 | .930 |
| 1000 | Lower | 0 | Uncorrected | 0.104 (0.001) | 1.05 | 1.05 | 1.17 | .068 | .068 | .114 |
| 1000 | Lower | 0 | Croon-DML | 0.016 (0.002) | 1.02 | 1.03 | 1.14 | .934 | .934 | .942 |
| 1000 | Lower | 0 | Croon-DML-R | 0.008 (0.002) | 1.02 | 1.03 | 1.13 | .944 | .944 | .948 |
| 1000 | Lower | 0.30 | Uncorrected | 0.103 (0.001) | 1.02 | 1.02 | 1.15 | .060 | .062 | .134 |
| 1000 | Lower | 0.30 | Croon-DML | 0.016 (0.002) | 1.01 | 1.02 | 1.13 | .926 | .928 | .936 |
| 1000 | Lower | 0.30 | Croon-DML-R | 0.009 (0.002) | 1.01 | 1.02 | 1.13 | .942 | .942 | .940 |
| 1000 | Higher | 0 | Uncorrected | 0.051 (0.001) | 1.00 | 1.00 | 1.13 | .614 | .614 | .676 |

| $N$ | Measurement quality | $\theta_0$ | Method | Bias (MCSE) | Conditional SE/SD | Stacked SE/SD | Boot SE/SD | Wald coverage (conditional) | Wald coverage (stacked) | BC coverage |
|---|---|---|---|---|---|---|---|---|---|---|
| 1000 | Higher | 0 | Croon-DML | 0.012 (0.001) | 0.99 | 0.99 | 1.11 | .930 | .928 | .940 |
| 1000 | Higher | 0 | Croon-DML-R | 0.010 (0.001) | 0.99 | 0.99 | 1.11 | .942 | .942 | .944 |
| 1000 | Higher | 0.30 | Uncorrected | 0.050 (0.001) | 1.04 | 1.04 | 1.17 | .642 | .640 | .676 |
| 1000 | Higher | 0.30 | Croon-DML | 0.011 (0.001) | 1.03 | 1.04 | 1.15 | .950 | .954 | .958 |
| 1000 | Higher | 0.30 | Croon-DML-R | 0.008 (0.001) | 1.03 | 1.04 | 1.15 | .954 | .954 | .956 |
| 5000 | Lower | 0 | Uncorrected | 0.117 (0.001) | 0.95 | 0.95 | 1.06 | .000 | .000 | .000 |
| 5000 | Lower | 0 | Croon-DML | 0.009 (0.001) | 0.95 | 0.95 | 1.06 | .896 | .898 | .934 |
| 5000 | Lower | 0 | Croon-DML-R | 0.002 (0.001) | 0.95 | 0.95 | 1.05 | .936 | .938 | .938 |
| 5000 | Lower | 0.30 | Uncorrected | 0.116 (0.001) | 1.01 | 1.01 | 1.14 | .000 | .000 | .000 |
| 5000 | Lower | 0.30 | Croon-DML | 0.009 (0.001) | 1.03 | 1.04 | 1.16 | .936 | .938 | .948 |
| 5000 | Lower | 0.30 | Croon-DML-R | 0.002 (0.001) | 1.03 | 1.04 | 1.16 | .954 | .956 | .954 |
| 5000 | Higher | 0 | Uncorrected | 0.054 (0.001) | 1.02 | 1.02 | 1.15 | .034 | .034 | .048 |
| 5000 | Higher | 0 | Croon-DML | 0.004 (0.001) | 1.01 | 1.01 | 1.13 | .944 | .944 | .946 |
| 5000 | Higher | 0 | Croon-DML-R | 0.002 (0.001) | 1.00 | 1.01 | 1.13 | .950 | .950 | .944 |
| 5000 | Higher | 0.30 | Uncorrected | 0.054 (0.001) | 0.99 | 0.99 | 1.12 | .030 | .030 | .042 |
| 5000 | Higher | 0.30 | Croon-DML | 0.005 (0.001) | 0.99 | 0.99 | 1.11 | .938 | .938 | .946 |
| 5000 | Higher | 0.30 | Croon-DML-R | 0.002 (0.001) | 0.99 | 0.99 | 1.11 | .940 | .940 | .954 |

*Note.* Uncorrected = uncorrected factor-score DML; Croon-DML-R = residualization-inflation correction; MCSE = Monte Carlo standard error; Conditional SE/SD and Stacked SE/SD = mean conditional and stacked sandwich standard error divided by the empirical standard deviation; Boot SE/SD = mean bootstrap standard error of Table S18 divided by the same SD; Wald coverage = proportion of 95% Wald intervals from the respective SE containing $\theta_0$; BC coverage = coverage of the bias-corrected bootstrap interval of Table S18 on the same samples. The samples, folds, learner, and point estimates are those of Table S18; the stacked SE treats the

smoother weights and the selected penalties as fixed (Section S1.4). The Wald coverages use all 500 replications; the BC coverage of uncorrected DML, Croon-DML, and Croon-DML-R at $N = 500$ with lower measurement quality is based on the 493 replications with bootstrap intervals (Table S18). With 500 replications, the MCSE of coverage is about 0.010. LOOCV = leave-one-out cross-validation; DML = double/debiased machine learning.

**References for the Supporting Information**